\documentclass[]{spie}  %>>> use for US letter paper
\usepackage{amsmath,amsfonts,amssymb}
\usepackage{graphicx}
\usepackage[colorlinks=true, allcolors=blue]{hyperref}

\title{The Infrared Imaging Spectrograph (IRIS) for TMT: closed-loop adaptive optics while dithering}

\author[a]{Edward L. Chapin$^*$}
\author[a]{Jennifer Dunn}
\author[a]{David Andersen}
\author[a]{Glen Herriot}
\author[a]{Dan Kerley}
\author[b]{Takashi Nakamoto}
\author[c]{Jimmy Johnson}
\author[c]{Lianqi Wang}
\author[c]{Gelys Trancho}
\author[c]{Eric Chisholm}
\author[c]{Brent Ellerbroek}
\author[c]{Kim Gillies}
\author[b]{Yutaka Hayano}
\author[d]{James Larkin}
\author[a]{Luc Simard}
\author[c]{Mark Sirota}
\author[b]{Ryuji Suzuki}
\author[e]{Bob Weber}
\author[f]{Shelley Wright}
\author[g]{Kai Zhang}

\affil[a]{National Research Council Herzberg, 5071 W Saanich Rd, Victoria, V9E 2E7, Canada}
\affil[b]{National Astronomical Observatory of Japan, 2-21-1 Osawa, Mitaka, Tokyo, 181-8588, Japan}
\affil[c]{Thirty Meter Telescope International Observatory, 100 W Walnut St, \#300, Pasadena, CA 91124, USA}
\affil[d]{Department of Physics and Astronomy, Univ. of California, Los Angeles, CA 90095-1547, USA}
\affil[e]{Caltech Optical Observatories,1200 E California Blvd., Pasadena, CA 91125, USA}
\affil[f]{Center for Astrophysics and Space Sciences, Univ. of California, San Diego, La Jolla, CA 92093, USA}
\affil[g]{National Astronomical Observatories of China-Nanjing Institute of Astronomical Optics and Technology, Nanjing, China}

\authorinfo{$^*$ed.chapin@nrc-cnrc.gc.ca, phone +1 250 363 6971}

\begin{document}
\maketitle

\begin{abstract}
 The InfraRed Imaging Spectrograph (IRIS) is the first-light client instrument for the Narrow Field Infrared Adaptive Optics System (NFIRAOS) on the Thirty Meter Telescope (TMT). IRIS includes three natural guide star (NGS) On-Instrument Wavefront Sensors (OIWFS) to measure tip/tilt and focus errors in the instrument focal plane. NFIRAOS also has an internal natural guide star wavefront sensor, and IRIS and NFIRAOS must precisely coordinate the motions of their wavefront sensor positioners to track the locations of NGSs while the telescope is dithering (offsetting the telescope to cover more area), to avoid a costly re-acquisition time penalty. First, we present an overview of the sequencing strategy for all of the involved subsystems. We then predict the motion of the telescope during dithers based on finite-element models provided by TMT, and finally analyze latency and jitter issues affecting the propagation of position demands from the telescope control system to individual motor controllers.
\end{abstract}

\keywords{coordinated motion control, real-time control, adaptive optics, wavefront sensors, TMT, IRIS, NFIRAOS}

\section{INTRODUCTION}
\label{sec:intro}  % \label{} allows reference to this section

The InfraRed Imaging Spectrograph (IRIS) will be the first-light workhorse instrument for the Thirty Meter Telescope (TMT). IRIS is fed by the Narrow Field Infrared Adaptive Optics System (NFIRAOS), which will provide diffraction-limited optical performance in J, H, and K bands, across the full 35\,arcsec$\times$35\,arcsec field-of-view (FOV) of the IRIS imager. Pickoff optics near the center of the IRIS FOV feed an integral field spectrograph (IFS), which can be run in parallel with the imager. NFIRAOS is a Laser Guide Star (LGS), Multi-conjugate Adaptive Optics System (MCAO) which, in its primary observing mode, uses an asterism of six artificial sodium laser guide stars [observed with corresponding Laser Guide Star Wavefront Sensors (LGS WFS)] to drive two deformable mirrors (DMs), conjugated to 0\,km (DM0) and 11.8\,km (DM11), at a rate of up to 800\,Hz. An additional natural guide star wavefront sensor (NGS WFS) within NFIRAOS provides reference ``truth'' measurements to compensate errors in the LGS system on longer timescales (e.g., radial Zernike error introduced by variations in the structure of the sodium layer). Light is steered on to the NGS WFS using a Star Selection Mechanism (SSM) which combines a movable X-Y stage, and a separate tip/tilt/focus stage. Low-order wavefront error measurements are provided by positionable natural guide star (NGS) On-Instrument Wavefront Sensors (OIWFS) and On-Detector Guide Windows (ODGW) within IRIS. These OIWFSs and ODGWs will be used to measure tip, tilt, and focus errors to which the LGS WFS is blind. NFIRAOS and IRIS are required to deliver 50\% sky coverage at the North Galactic Pole. In this context sky coverage refers to the ability of the OIWFS and ODGW to provide sufficient information to control the low-order modes and meet the TMT LGS MCAO wavefront error (WFE) budget. The OIWFS and ODGW will also control the six plate scale modes at the science focal plane.
In addition to the DMs, NFIRAOS can use several slower mechanisms to compensate for longer timescale, and larger amplitude errors. DM0 is mounted to a tip/tilt stage (TTS) with a control bandwidth of approximately 20\,Hz, and acts as a ``woofer'' to compensate slow and large-amplitude tip/tilt errors, while the DM itself acts as a ``tweeter'' for the higher-frequency and lower-amplitude errors. Longer timescale pointing and tip/tilt errors are offloaded to the TMT Telescope Control System (TCS) for compensation.

\begin{figure}[ht]
\begin{center}
\includegraphics[width=\linewidth]{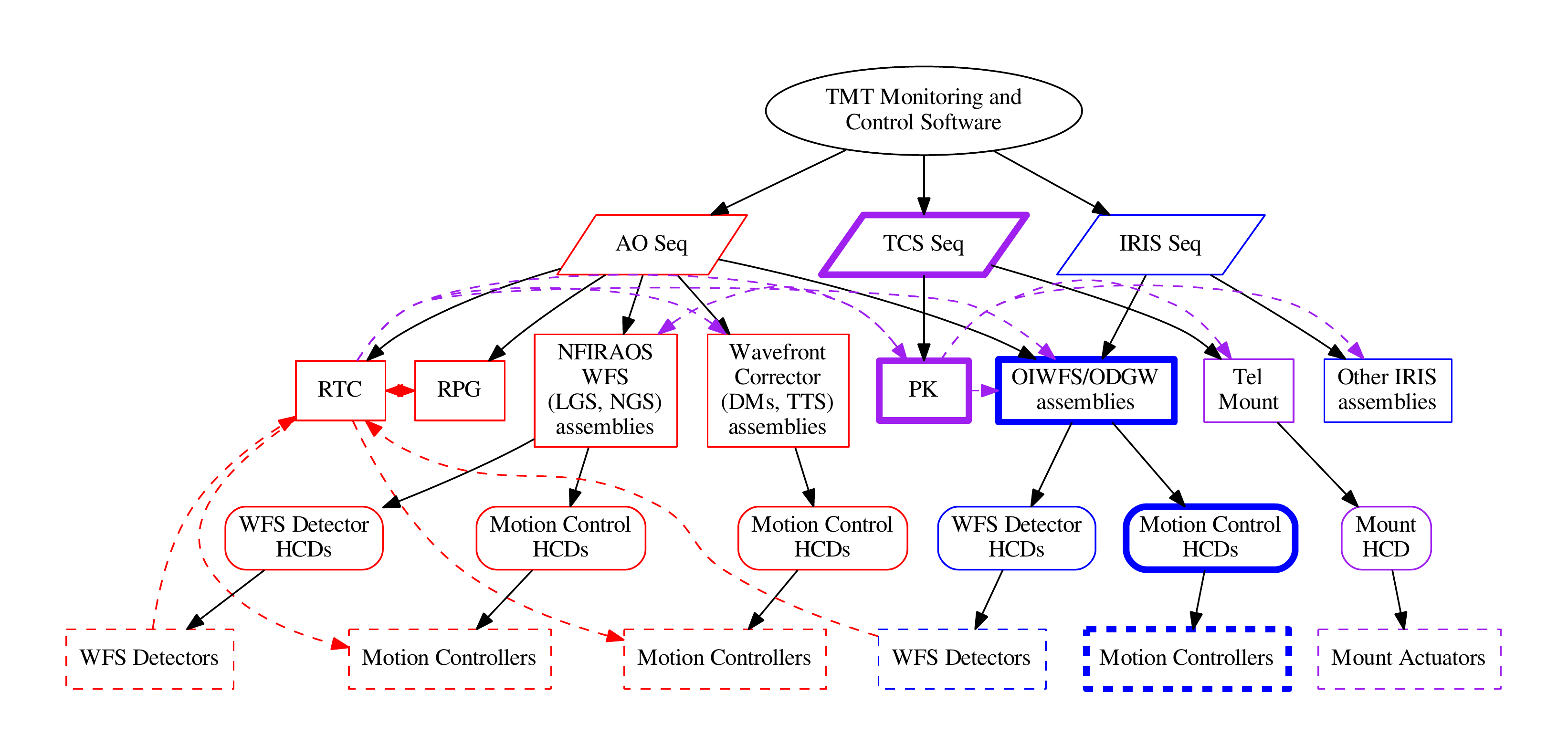}
\end{center}
\caption{\label{fig:hierarchy}
A representative portion of the observatory software hierarchy relevant to AO operations using IRIS and NFIRAOS. From top to bottom, the layers of this hierarchy are referred to as: the Monitoring and Control Layer; the Sequencing Layer; the Assembly Layer; the Hardware Control Layer; and finally the physical Hardware Layer. Subsystems are color-coded: red for AO; blue for IRIS; and purple for the TCS. Black lines indicate the configuration hierarchy (e.g., sequencers configure assemblies, and assemblies configure HCDs). Note that while the OIWFS/ODGW assemblies are part of IRIS, they are primarily configured by the AOSq. Event Service communication is shown with purple dashed lines; for example, the PK publishes demands for various assemblies within NFIRAOS, IRIS, and the telescope mount, and the RTC publishes corrections for many of the same positioners. A separate high-speed network and protocol, depicted with red dashed lines, is used by the RTC to receive data directly from the WFS, send commands to wavefront correctors, and exchange data with the RPG. The components highlighted in bold are included in the prototype described later in this paper.}
\end{figure}

Observing with NFIRAOS and IRIS is a complex task, involving communication and feedback between a number of telescope subsystems at different rates, as shown in Fig.~\ref{fig:hierarchy}. Within NFIRAOS is a Real Time Controller (RTC) that is responsible for processing measurements from all of the wavefront sensors (WFS), and providing demands to all of the above mentioned wavefront correctors up to the peak 800\,Hz rate. Statistics collected by the RTC are sent to the Reconstructor Parameter Generator (RPG) which periodically revises and uploads ``control matrices'' to the RTC, which it in turn uses as part of its real-time wavefront corrector calculations. The communication between WFSs, wavefront correctors, the RTC and the RPG use a dedicated high-bandwidth, low-latency network and protocol, and highly-optimized software written in C to meet the stringent real-time requirements. The NFIRAOS RTC is discussed elsewhere in these proceedings\cite{dunn2018}.

In contrast, the NGS WFS, OIWFSs, and ODGWs are positioned using continuous demand streams that are sent by the TCS, which converts star coordinates into focal plane coordinates using its pointing kernel (PK), and with error correction streams optionally provided by the RTC. The TCS also provides position streams for numerous other devices throughout the observatory, including the telescope mount. These streams are typically provided at rates 20--100\,Hz, depending on the requirements of the device. These demands are sent using the TMT ``Event Service''.

Overall coordination and configuration of the various subsystems at TMT will be orchestrated by ``sequencers'', as depicted in Fig.~\ref{fig:hierarchy}. NFIRAOS and the RTC, as part of the TMT Adaptive Optics (AO) system, will be configured by the Adaptive Optics Sequencer (AOSq). The TCS and science instruments will also have their own sequencers. The sequencers convert requests from higher-level subsystems (e.g., ``configure the AO and science instruments for a particular observation'', ``slew to the location of the observation'', ``engage the AO system'', and finally ``commence science exposures'') into sequences of commands relayed by the TMT Command Service to particular components within each subsystem for which the sequencer is responsible (e.g., configure individual WFS detectors, slew the telescope mount, move filter wheels). The software components with which sequencers interact are called assemblies, with each assembly typically exposing a simple unified interface for a single device (e.g., a detector, filter wheel, beam splitter etc.). Finally, assemblies interact with Hardware Control Daemons (HCD) that directly interface with physical hardware. For example, an HCD for a motor controller may accept commands from an assembly to drive a stage a particular number of millimeters, and then convert the request into motor counts and forward the request to the controller using its particular communications protocol.

All of the TMT software systems mentioned above are being developed in parallel, but with staggered schedules. TMT Common Software\cite{gillies2016} (CSW), which includes the Event and Command Services, as well as a framework for building software components such as assemblies and HCDs, passed its final design review in Jan. 2017, and a production version is currently being developed by an India-based vendor named ThoughtWorks, under contract to the India TMT Co-ordination Centre (ITCC). Software for NFIRAOS (being developed at NRC Herzberg), including the RTC, and slow opto-mechanical mechanisms, such as the tip/tilt stage, and NGS WFS mentioned above, recently passed their Final Design Reviews. AO Executive Software (AOESW), is being developed at TMT; the AOSq is in the Preliminary Design Phase, and the RPG the Final Design Phase. IRIS software (led by NRC Herzberg, with U.S. and Japanese partners) has just commenced its Final Design Phase. The TCS [developed at TMT, primarily by the ITCC with support of the Inter-University Centre for Astronomy and Astrophysics (IUCAA)] is currently in its Preliminary Design Phase. Finally, the remaining observatory software work packages will commence their Preliminary Design Phases this autumn.

Early prototyping efforts that involve multiple subsystems are therefore critical to gain confidence in the overall observatory software architecture. To this end, this paper describes motion analysis and simulations that have informed the designs of NFIRAOS and IRIS, making use of simulations and prototype software provided by TMT. In particular, we examine one of the more challenging observational scenarios that will be faced by the slow opto-mechanical mechanisms within NFIRAOS and IRIS: closed-loop dithering.  In order to cover large areas of sky efficiently, TMT will sequentially offset up to 30 arcsec on-sky, performing separate science integrations at each location, as part of a single observing sequence. The NGS WFS and OIWFS positioners must undergo complementary motion to that of the telescope during these offsets so that their guide stars can be observed continuously. If the WFSs lose lock on their guide stars, NFIRAOS will cease performing AO corrections, and a substantial AO re-acquisition time penalty will be incurred upon completion of the dither offset (potentially as much as tens of seconds). Such a penalty would render dithering (and therefore large-area mapping) incredibly inefficient.

In Section~\ref{sec:motion} the expected motion of the telescope during a dither is analyzed. The results of this analysis are used to justify a TCS demand publication rate 20\,Hz for the control of NFIRAOS and IRIS WFS positioners. Then, Section~\ref{sec:swproto} uses CSW framework and service prototypes [which run on Java Virtual Machines (JVM)], to create a ``vertical slice'' through the control system, including the TCS, an IRIS assembly, an HCD, and a real motion controller. Timing measurements are used to explore latency and jitter in the control system, ultimately demonstrating that the motion requirements can be met.

%The , a publish-subscribe system based on the Redis key-value store.

%Prototype use of CSW services: JVM, Akka, Reddis

%Expected motion of telescope.

\section{Telescope dither motion}
\label{sec:motion}

A dither consists of the telescope moving smoothly between two locations on the sky in order to place the target at a different location in the science instrument focal plane (in this case, IRIS). During this time the AO loops should remain closed to minimize lost observing time associated with re-acquiring WFS targets, meaning that any natural guide star wavefront sensor positioners (the NGS WFS SSM in NFIRAOS, and the IRIS OIWFS movable probes) must undergo precise complementary motion to that of the telescope. NFIRAOS is able to compensate for some errors in this motion; the RTC (which continually receives data from the WFSs) can sense tip/tilt drift, and attempts to remove it using a combination of tip/tilt stage (TTS) motion (for slower, large-amplitude drift) and the DM (faster, low-amplitude drift). However, if the low-frequency error exceeds the stroke of the TTS available for such corrections (or if the control system is unable to respond in a timely manner), the WFSs may lose lock. Similarly, excess high-frequency tip/tilt errors during the dither may exceed the dynamic range of the WFS, potentially resulting in unacceptable image blur and/or loss of lock for the WFSs.

\subsection{Approaches to WFS position error mitigation}
\label{sec:approaches}

There are several mechanisms that might alleviate the problem of excess low-frequency noise. First, if the amplitude of the errors lies within the field-of-view (FOV) of the WFSs (i.e., they are able to continue taking measurements), an anti-windup filter may be incorporated into the RTC to ensure that the TTS is not saturated (a non-trivial addition to an already complex Kalman filter in the current design). Second, if the errors exceed the FOVs of the WFSs the RTC could send the error streams to the WFS positioners so that they ``self-guide''. For reference, early designs of the OIWFS were based on Teledyne HAWAII-1RG detectors that would have been restricted to small regions-of-interest (at the frame rates required throughout dithers). Recently, it was decided to update the design to use near-infrared Avalanche Photodiode (APD) detectors, resulting in a FOV approximately 1.5\,arcsec that can be read at full frame rate; while the full impact of this choice has yet to be explored, it will help mitigate the risk of losing lock.

For the purposes of this study, we proceed with what seems like a simpler third option: driving the telescope in a way that minimizes position errors. Adopting saturation of the TTS as the limiting error at low frequencies, we seek a design whereby the OIWFS probe and NGS WFS positioners are commanded by the TCS with an accuracy of 0.2\,mm (or 0.1\,arcsec) in the NFIRAOS input focal plane (noting that the platescale is approximately 2\,mm/arcsec) continuously throughout the dither. It is assumed that high-frequency errors that might cause image blur are either negligible or can be removed by the TTS and DM in this study.

\subsection{Telescope mount simulations}

\begin{figure}[ht]
\begin{center}
\includegraphics[width=0.8\linewidth]{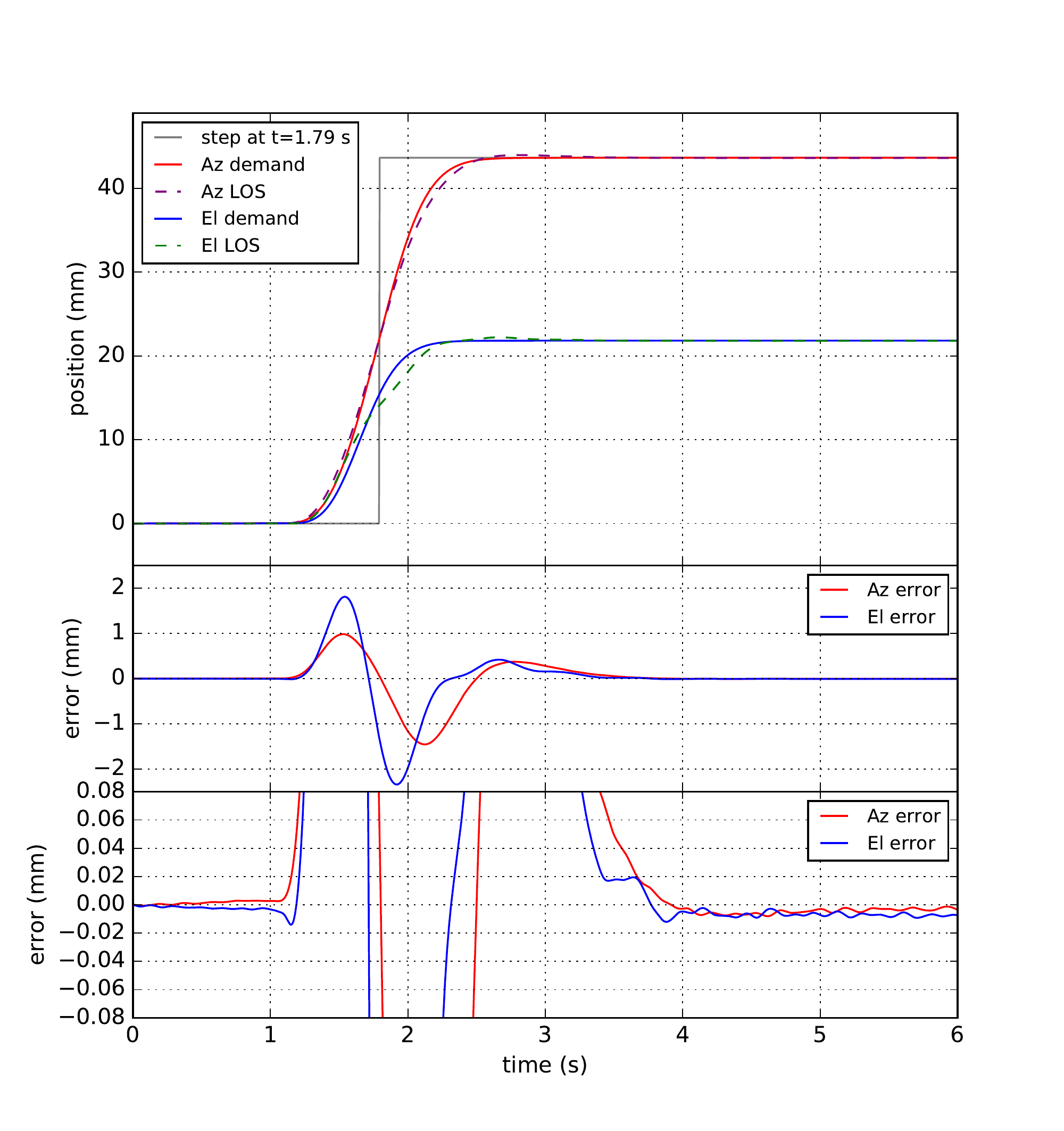}
\end{center}
\caption{\label{fig:fe_pos}
Representative output of the MELCO finite element model for a short telescope offset covering 14\,arcsec on-sky (20\,arcsec in azimuth, 10\,arcsec in elevation, at an elevation of 60\,deg), or $\sim$30\,mm in the NFIRAOS focal plane. The demands and resulting line-of-sight values are shown in the top plot (with an instantaneous step function for reference), the middle plot shows the full amplitude of the errors, and the bottom plot shows the errors with a smaller vertical range to enhance the view of the settling behaviour at $t\gtrsim3$\,s.}
\end{figure}

In 2016 TMT commissioned a trade study by MELCO to analyze the telescope mount control system and structural response to different types of motion. The basis of their analysis was a finite element (FE) model of the telescope structure, and the mount control system that included, among other things, a trajectory waveform generator, and realistic feedback mechanisms (e.g., from elevation and azimuth encoders). The most pertinent scenarios examined in that study were a series of offsets of comparable amplitude to the dithers mentioned above (e.g., covering 10s of arcsec on-sky), though completing in significantly shorter time intervals (of order $\sim$2--3\,s, vs. the 5\,s being considered for NFIRAOS and client instruments). These fast offsets were a primary focus of the trade study to ensure that the telescope could meet top-level settling time requirements that are driven by seeing-limited observing modes.

An example simulation is shown in Fig.~\ref{fig:fe_pos}. The modeled motion exhibits a significant error signal [comparing the telescope line-of-sight (LOS) with the mount demands], and transient ringing upon its completion. In this case, the errors are as large as $\sim$1--2\,mm in both azimuth and elevation in the middle of the move (clearly exceeding the 0.2\,mm goal set in the previous section). For the purpose of closed-loop AO dithering described in this paper, such motion is not ideal as the TTS and WFS positioners would need to track that motion to a high-degree of accuracy, including large accelerations and velocities, while accounting for the error to avoid losing lock with the WFSs. The error mostly subsides by $t\sim4$\,s, though higher-frequency ringing persists (amplitude $\sim0.0025$\,mm) past the end of the plot, presenting a separate challenge to the control system.

\subsection{Smooth ``S'' curve motion}

\begin{figure}[ht]
\begin{center}
\includegraphics[width=0.8\linewidth]{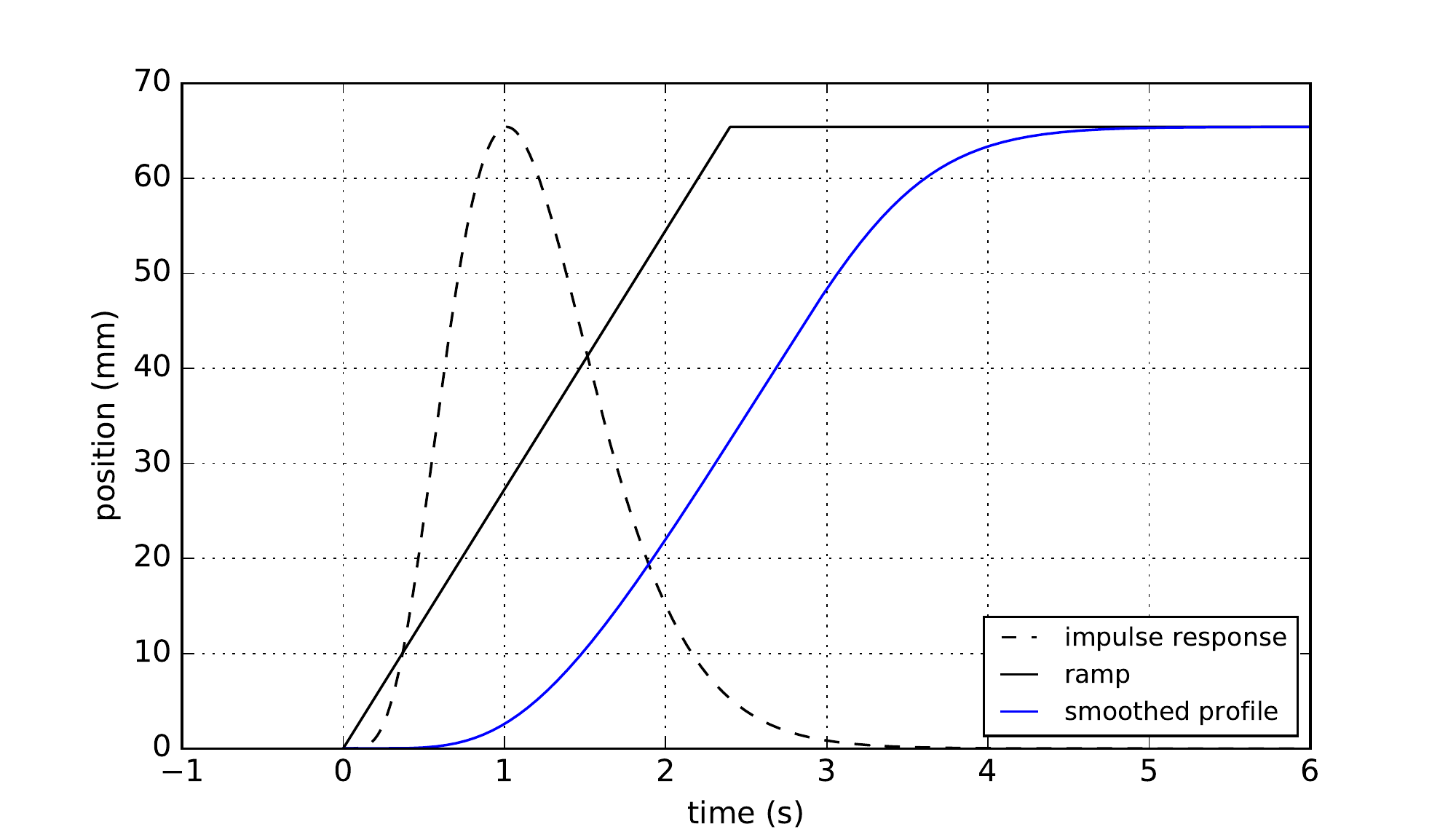}
\end{center}
\caption{\label{fig:s_curve}
A maximal dither profile used to command the telescope mount and WFS positioners, covering 30 arcsec (65.4\,mm in the NFIRAOS focal plane), over an interval of 5\,s. The smoothed profile (blue line) is constructed by convolving a linear ramp (black line) with a 6th-order anti-jerk filter (dashed line).}
\end{figure}

In order to minimize pointing mismatch, we consider a gentler motion profile that has smooth acceleration and higher-order derivatives, and occurs over a longer interval. The proposed waveform is depicted in Fig.~\ref{fig:s_curve}. It implements the most challenging dither that NFIRAOS and IRIS are required to accommodate: travel over a distance of 30\,arcsec (or 65.4\,mm in the focal plane; a substantial fraction of the IRIS 50-arcsec FOV) in a time of 5\,s.

This ``S'' curve was constructed by convolving a linear ramp, $r(t)$, with a 6th-order ``anti-jerk'' filter, $f(t)$, 
\begin{eqnarray}
  f(t) &=& (n-1)!t^{n-1}e^{-at} \\
  r(t) &=& \begin{cases}
    d \frac{t}{t_r} &, t < t_r \\
    d               &, t \ge t_r
   \end{cases} ,
\end{eqnarray}
with $n=6$, $a=4.93$, $d=65.4$\,mm, and $t_r=2.4$\,s.

In order to estimate the response of the telescope mount to this gentler waveform, we have bootstrapped the results of the telescope mount trade study in the following way. If we assume a linear model for the telescope response to arbitrary mount driving functions, this response can be inferred from the FE model using Fourier analysis. Expressing the telescope LOS direction (an output of the FE model), $\mathrm{LOS}(t)$, as the convolution of the driving demand function, $\mathrm{Demand}(t)$, and some impulse response function, $I(t)$, i.e., $\mathrm{LOS}(t) = \mathrm{conv}\{\mathrm{LOS}(t),I(t)\},$ we solve for $I(t)$ by performing Fourier de-convolution of the LOS and demand ``data'' from the simulations:
\begin{equation}
\label{eq:decon}
  I(t) = \mathrm{FFT}^{-1} \bigg\{
    \frac{\mathrm{FFT}\{\mathrm{LOS}(t)\}}
    {\mathrm{FFT}\{\mathrm{Demand}(t)\}}
    \bigg\} ,
\end{equation}

In practice, to suppress ringing effects caused by the implicit assumption of periodicity in the LOS and demand waveforms in Fourier analysis, the Demand and LOS time series are padded substantially before and after the moves occur, and Hanning window functions are applied. Finally, at higher frequencies ($\gtrsim 2$\,Hz, depending on the particular simulation) noise in the denominator of the right-hand side of Eq.~\ref{eq:decon} (in particular values approaching zero) causes the results to diverge; the positive tail of a Gaussian curve is thus fit to the data (separately in amplitude and in phase) in this regime so as to smoothly roll-off the power in $I(t)$ toward higher frequencies.

The procedure is applied separately to the simulated azimuth and elevation data, to derive independent azimuth and elevation impulse response functions. Once obtained, the demand waveforms are re-convolved with our measured $I(t)$ and compared with the LOS data output by the simulations. The padding, windowing, and Gaussian roll-off are then hand-tuned to ensure that residual (primarily high-frequency) errors are insignificant compared to the 0.2\,mm error threshold discussed in Section~\ref{sec:approaches}.

Finally, this model is tested for a range of simulations that were produced as part of the trade study: it was found that single azimuth and elevation impulse response functions are able to reproduce the resulting LOS curves over a range of telescope starting elevations and offset amplitudes, confirming that our assumption of linearity is valid on scales relevant to our analysis.

\begin{figure}[ht]
\begin{center}
\includegraphics[width=0.8\linewidth]{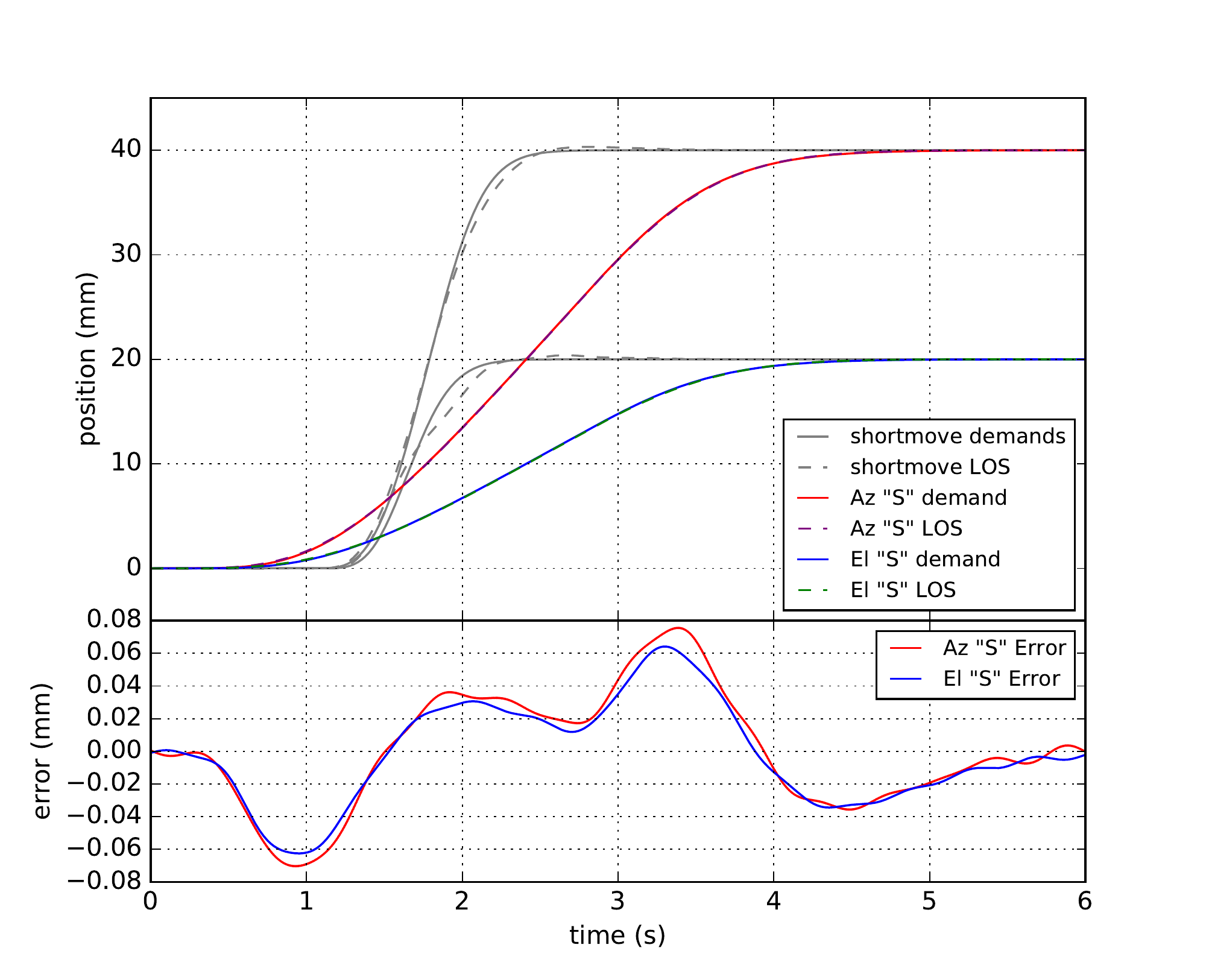}
\end{center}
\caption{\label{fig:antijerk_motion}
Comparison between the short telescope offset FE model simulation of Fig.~\ref{fig:fe_pos} [grey lines demands, dashed grey lines line-of-site (LOS) response, scaled by a factor of $\sim$2], with the demands (red and blue lines) and extrapolated LOS responses (purple and green dashed lines) expected from the gentler ``S'' curve proposed for closed-loop AO dithers. The errors for the ``S'' curve (bottom plot) are now well within the desired 0.2\,mm.
}
\end{figure}

%\begin{figure}[ht]
%\begin{center}
%\includegraphics[width=0.8\linewidth]{transfer_err_S1_c5b.pdf}
%\end{center}
%\caption{\label{fig:antijerk_error}
%Error signals (azimuth, elevation, and total absolute on-sky error) projected into the% instrument focal plane, showing that it does not exceed 0.2\,mm.
%}
%\end{figure}

With confidence in the impulse response functions, they are then applied to the ``S'' curve. A comparison between the original shorter moves from the FE model (scaled from 14\,arcsec to 30\,arcsec amplitude) to the filtered ``S'' curve is shown in Fig.~\ref{fig:antijerk_motion}. While the error in the shorter move is clearly visible, it is substantially smaller for the ``S'' curve as shown in the bottom plot of the figure, with a peak-to-peak error of $<$0.07\,mm (less than our 0.2\,mm requirement). The error plot also exhibits less high-frequency ringing, as one would expect given a driving function with less high-frequency power. However, the true high-frequency behaviour would need to be confirmed with an updated FE model since the inferred impulse response functions are truncated at $f\gtrsim2$\,Hz, as noted above.

%\textbf{more notes:}
%
%\begin{itemize}
%
%\item The error signal measured by mount encoders (a separate plot, not yet shown) could be fed into the WFS positioner demand streams to reduce the error by another factor of $\sim$10.

%\item haven't said anything about high-frequency error yet. In the Melco simulations, the high-pass filtered short move has a residual error of about 7 milli-arcsec peak-to-peak. It would obviously be a lot less for a gentler ``S'' curve, though the impulse response functions above are not good enough at high-frequency to demonstrate this. 

%\end{itemize}

\subsection{TCS demands}
\label{sec:tcs_demands}

In the current design the TCS will publish demand streams for the NFIRAOS and IRIS WFS positioners at a rate of 20\,Hz. Assuming the ``S'' motion profile of the previous sections, we assess the adequacy of this rate under some bounding assumptions about the overall control strategy:
\begin{itemize}

\item position demands are published by the TCS instantaneously (i.e., at the precise moment they are valid), along with a time stamp;

\item latency in the transfer of demands from the TCS to the positioner actuators is no longer than one 20\,Hz period, or 50\,ms; and

\item the software assemblies and/or HCDs controlling the actuators are able to buffer past demands, and extrapolate forward in time so as to estimate the correct demands at the instant they are to be sent to the relevant hardware controllers (which implies that all of the clocks are synchronized).

\end{itemize}

\begin{figure}[ht]
\begin{center}
\includegraphics[width=0.8\linewidth]{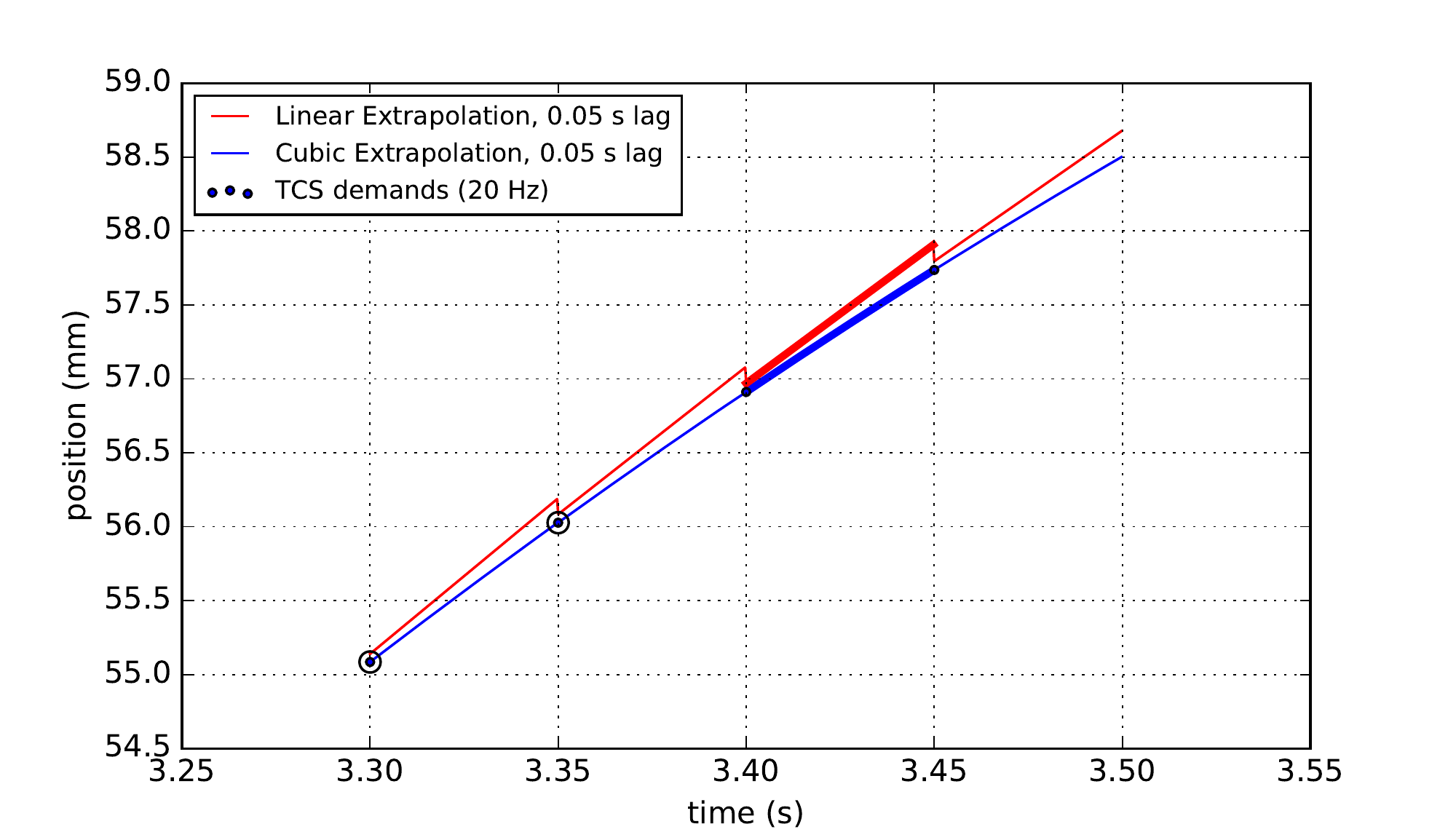}
\end{center}
\caption{\label{fig:dither_extrap}
Linear and cubic extrapolation of 20\,Hz TCS demand waypoints for WFS positioners, when the demands arrive 50\,ms late. For illustration, the bold line segments result from fits to the circled demands (additional earlier demands, not shown, were also used in the case of the cubic).
}
\end{figure}

\begin{figure}[ht]
\begin{center}
\includegraphics[width=0.8\linewidth]{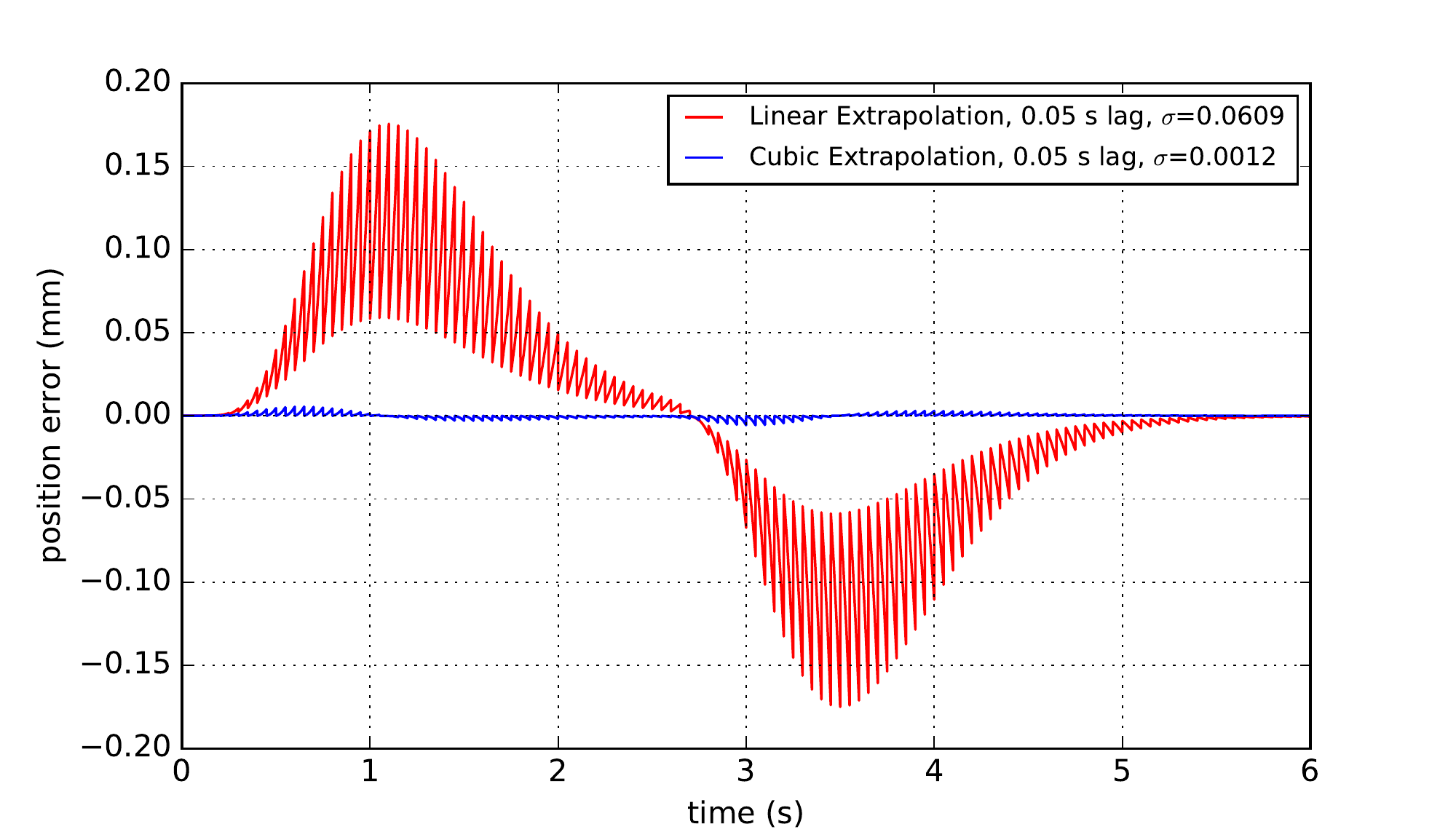}
\end{center}
\caption{\label{fig:dither_extrap_err}
Errors between linear- and cubic-extrapolated 20\,Hz TCS demands and the true position, when the demands arrive 50\,ms late.
}
\end{figure}

Fig.~\ref{fig:dither_extrap} shows a zoom-in of the ``S'' curve in Fig.~\ref{fig:s_curve} during a period of maximum acceleration ($t\sim3.4$\,s). Each dot represents a single sample of that curve at a rate of 20\,Hz. The red line indicates a linear extrapolation of the previous two TCS samples, applied over a period of 50\,ms, but delayed until 50\,ms after the demand was sent by the TCS (to simulate lag in the control system). Similarly, the blue line shows the result of a cubic spline extrapolation (this time to the most recent four samples received). For illustration, the circled TCS demands were used to fit the extrapolated line segments drawn in bold (note that an additional two earlier TCS demands, not shown, were also used in the case of the cubic).

Fig.~\ref{fig:dither_extrap_err} shows the error signal for both extrapolation strategies when compared to the ideal curve of Fig.~\ref{fig:s_curve}. The linear extrapolation case falls just within the desired error envelope of 0.2\,mm, while the cubic extrapolation results in better than an order-of-magnitude improvement.

We conclude that 20\,Hz is a sensible TCS demand publication rate, giving NFIRAOS and IRIS the option of a basic control strategy that uses linear fits to TCS position demands (i.e., constant-velocity). If this strategy proves to be inadequate, it can be mitigated using a higher-order fitting function.

\section{Vertical Slice Software Prototype}
\label{sec:swproto}

Making use of the CSW prototype framework and associated services that were developed as part of its Final Design Phase\footnote{https://github.com/tmtsoftware/csw}, a subset of the control system depicted in Fig.~\ref{fig:hierarchy} (components highlighted in bold) was implemented to assess the feasibility of the closed-loop AO dither scenario for the slow opto-mechanical mechanisms from a communications perspective. Specifically, it includes the TCS (to publish position demands), the IRIS Probe Arm Assembly (POA), and a Galil motion controller HCD. It does \textit{not} include initial sequencing details, such as high-level requests to the AOSq, IRIS, and TCS Sequencers to prime all of the involved mechanisms, and execute the AO acquisition sequence, which would result in many lower-level commands to assemblies and HCDs. The goal is to study the steady-state behaviour once the AO system has already been engaged.

\subsection{CSW overview}
\label{sec:csw}

The CSW framework and services are written in Scala\footnote{https://scala-lang.org}, a multi-paradigm (functional/object-oriented) language that targets the Java Virtual Machine (JVM). Concurrency is achieved using the Akka toolkit\footnote{https://akka.io}, an implementation of the ``actor model''. Put simply, applications are designed as collections of actors, which have a few basic properties: (i) actors communicate with each other strictly by passing immutable messages (i.e., there is no shared memory); (ii) actors handle one message at a time (i.e., they are single-threaded), and can continue to queue messages while handling a previous message; and (iii) actors can create other actors [actors belong to ``actor systems'', with the parent of all actors in a system referred to as the ``top level actor'' (TLA)]. In most cases, it is desirable to design actors such that they handle messages quickly, ensuring that they remain responsive. Applications that use CSW may be written either in Scala, or in Java (for which an API has also been supplied).

Actors can send messages both to other actors within the same actor system, other actor systems within the same JVM, and also actors running in different JVMs (including remote servers). Message passing between actors in the same JVM is equivalent to function calls, with the message itself being an object of any class. On the other hand, message passing between different JVMs (``remoting'') requires serialization and a transport layer (e.g., TCP or UDP sockets). While any object may also be used as a message in these cases (as long as a serializer is also provided), CSW serialization uses Protocol Buffers for commands and Java serialization for events (in the CSW prototype), and constrains each to be represented as sets of key-value pairs. Akka only guarantees that messages will be delivered at most once; it is up to the developer to add handshaking if confirmation of delivery is required (i.e., messages could go missing over the network when remoting).

The Command Service is a thin wrapper over Akka messaging. As mentioned above, a command is essentially a set of key-value pairs that are sent from one software component to another, though restricted to the hierarchy depicted in Fig.~\ref{fig:hierarchy} (e.g., an assembly may only received commands from a sequencer; it cannot receive commands from another assembly). CSW provides several mechanisms for accepting, tracking and reporting command completion to senders, though it is only used in its simplest ``fire and forget'' configuration in this work.

The Event Service is the TMT publish-subscribe system. It can be used, among other things, to send continuously changing telemetry streams (e.g., the TCS providing the locations of WFS positioners and the telescope mount), or for an assembly or HCD to report state changes (to which multiple entities may subscribe). Again, the contents of events are restricted to key-value pairs, with the underlying broadcast of published values to all subscribers accomplished through the use of Redis\footnote{https://redis.io}. Unlike commands, events may be passed between components at any level of the command hierarchy (e.g., one assembly may publish events used by another assembly).

Assemblies and HCDs typically extend actor base classes provided by the CSW framework. The framework is also responsible for starting and stopping the components, once it has been defined how they will be instantiated (i.e., whether they will all run in the same JVM, or in multiple JVM instances). The CSW Location Service is fundamental to the operation of systems of software components. Assemblies and HCDs register themselves with the Location Service upon startup, and query the Location Service for connection details of other software components and services.

\subsection{Software design}

\begin{figure}[ht]
\begin{center}
\includegraphics[width=0.6\linewidth]{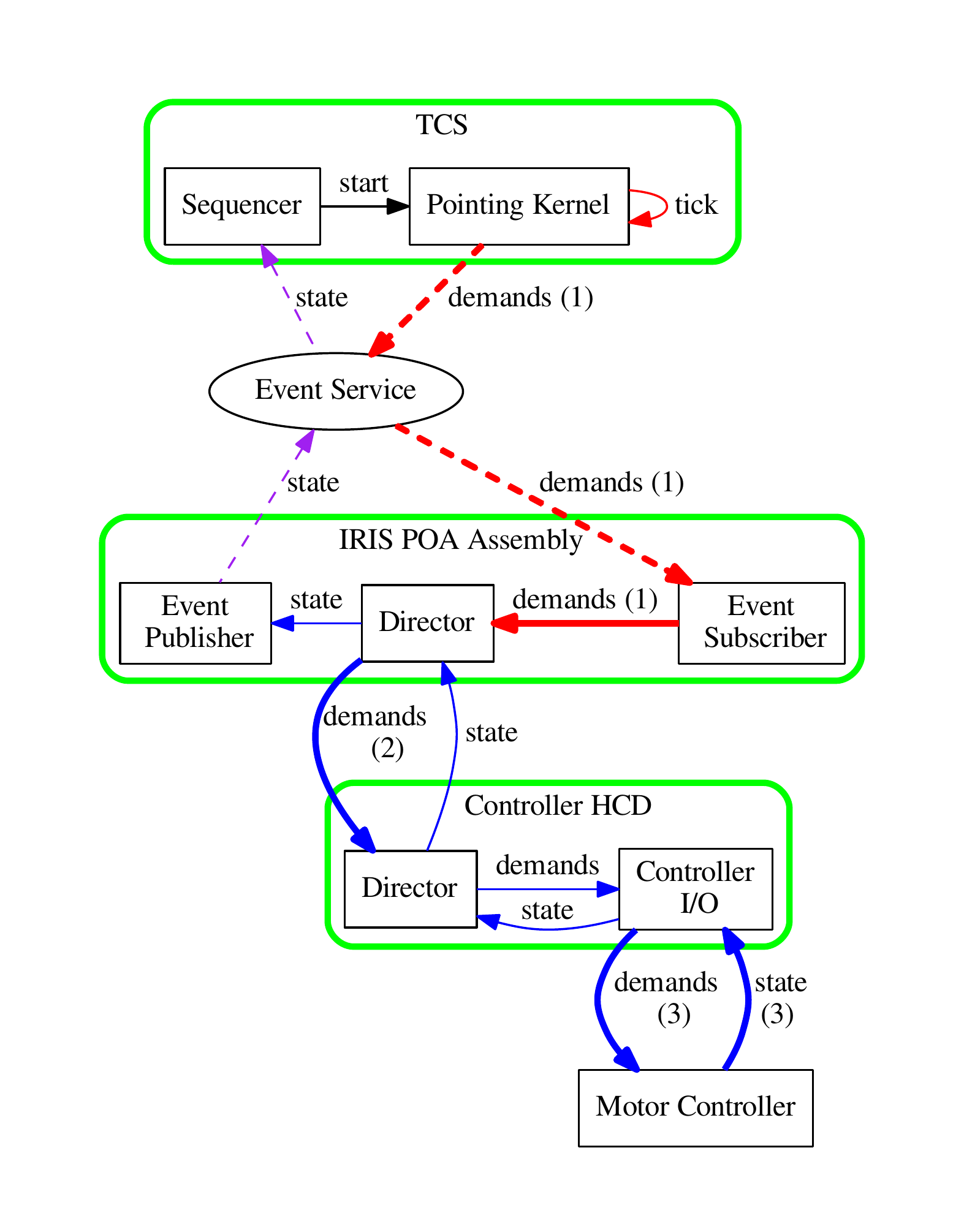}
\end{center}
\caption{\label{fig:actors}
Software decomposed into separate applications/JVM instances (green boxes), and actors (black boxes). Akka peer-to-peer messages (both within an application, or between applications using the CSW Command Service), and the socket communications with the Motor Controller, are depicted with solid lines, while messages transported via Redis as part of the CSW Event Service are shown with dashed lines. Messages that are triggered in a sequence are given the same color. Finally, intervals that are used for timing measurements are indicated with bold lines and a number [(1) TCS demand event delivery; (2) POA commands to HCD; (3) HCD controller I/O loop period].}
\end{figure}

\begin{figure}[htp]
\begin{center}
\includegraphics[width=0.8\linewidth]{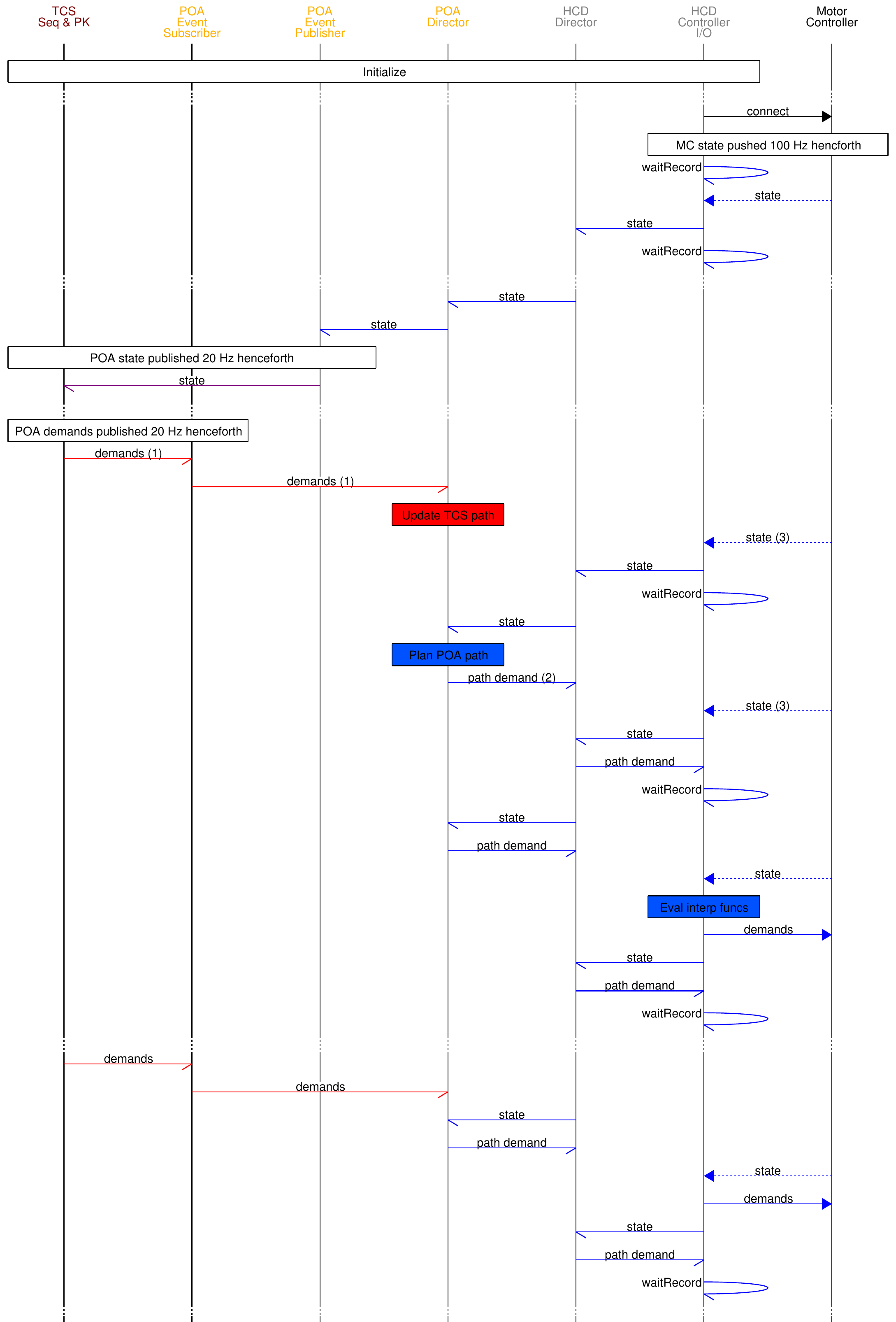}
\end{center}
\caption{\label{fig:seq}
Actor message passing sequence. Actors from the same application have the same color (to save space the TCS Sequencer and Pointing Kernel actors are depicted as one). Message sequences are also color-coded, and numbered timing intervals indicated to match Fig.~\ref{fig:actors}.}
\end{figure}

The software is designed as three separate applications: the TCS, POA Assembly, and Controller HCD. At the telescope the TCS and CSW services would likely all run on separate servers, while the POA assembly and HCD would typically run on one server (the IRIS OIWFS ``Components Controller''). However, the tests described here all run on a single server due to the availability of hardware at the time of writing. Each application is composed of a number of actors, as shown in Fig.~\ref{fig:actors}. The message passing sequence between the actors is provided in Fig.~\ref{fig:seq}. Each of the applications is run in a separate JVM instance (i.e., any messages passed between the applications will require serialization as described in Section~\ref{sec:csw}).

The TCS is the simplest application to understand. The \textbf{Sequencer} actor subscribes to state updates from the POA assembly via the Event Service. Upon reception of the first update (which gets transported by Redis, and then converted into an Akka message by the CSW Event Service client API), it immediately sends a local \texttt{start} message to the \textbf{Pointing Kernel} actor, to commence continuous publication of POA position \texttt{demands}. Publication is triggered via a \texttt{tick} message that is sent to itself on a fixed schedule, at a rate of 20\,Hz. The sequence of red messages in Fig.~\ref{fig:actors} are triggered by each of these \texttt{tick} messages. The waveform that it publishes is simply the ``S'' curve evaluated at 20\,Hz, repeated as three sets of ordered $(x,y)$ pairs, one for each of the OIWFS probes. In practice, the real TCS will be composed of several separate applications, rather than combining them as shown here.

The POA assembly involves three actors. An \textbf{Event Subscriber} makes use of an Event Service client to await \texttt{demands}. Once received, they are translated into local Akka messages and forwarded to the \textbf{Director} actor. This actor is the designated ``keeper of state''. It monitors demands from the TCS and stores them in local state variables. It also receives continuous \texttt{state} updates from the HCD via the CSW Command Service. In this way, the \textbf{Director} has all of the most recent information required to implement its inter-probe collision avoidance strategy\cite{chapin2016}. In the real system, it would execute one iteration of the path planning algorithm each time it receives an HCD state update, and respond by sending a new set of demands to the HCD. However, in the current test application it simply passes the most recent TCS demands on to the HCD verbatim. It also sends a state update message to the \textbf{Event Publisher}, which publishes state updates as events, though rate-limited to 20\,Hz. The message sequence that is driven by HCD updates is coloured blue, and runs at 100\,Hz. The rate-limited state events are published at 20\,Hz (independently of the rate at which demands are sent by the TCS).

Finally, the Controller HCD application consists of two actors. The \textbf{Controller I/O} actor performs all low-level communication with a real Galil DMC 4080 8-axis motion controller\footnote{http://www.galilmc.com/motion-controllers/multi-axis/dmc-40x0} over Ethernet using the vendor-supplied \texttt{gclib} via a Java Native Access (JNA) wrapper\footnote{http://galil.com/sw/pub/all/doc/gclib/html/java.html}. The controller is configured to asynchronously push state records, and a blocking call \texttt{GRecord()}\footnote{The authors extended the Galil JNA wrapper to include support for the \texttt{GRecord()} call, which was missing.} is used to retrieve them. Reception of the state record triggers the transmission of a state update as a normal Akka message to the HCD \textbf{director}, which in turn forwards the state to the assembly via the Command Service. The \textbf{Controller I/O actor} then immediately generates and sends new axis demands to the controller by evaluating the most recently received trajectories (parameterized as lines), evaluated at the precise instant that they will be sent. This procedure implements the extrapolation strategy described in Section~\ref{sec:tcs_demands}. Finally, the \textbf{Controller I/O} actor sends itself a message \texttt{WaitRecord} so that it begins waiting for the next state record. While \textbf{Controller I/O} is handling this control-loop message, in parallel, the assembly will have responded to the state update from the HCD \textbf{Director} by sending the next set of demands (updated trajectory expressed as a line) to the HCD \textbf{Director}, which then forwards them to the \textbf{Controller I/O} actor. Typically the \textbf{Controller I/O} actor will be handling the blocking wait for a state record, meaning that the trajectory update will not be received until the next iteration of the controller read/write loop.

While the \textbf{Controller I/O} actor sends mock position demands to the real controller for timing purposes, a basic simulation is needed to calculate the positions of each axis at each iteration of the controller read/write loop. Each time a new demand is calculated, each of the probes move their two stages (rotational and linear extension\cite{chapin2016}) up to their maximum rates toward the new targets, with velocity changes being instantaneous. A more realistic simulation might include a trapezoidal velocity profile (i.e., including constant acceleration), and a PID loop. Such an improvement is a goal for future work.

The choice to set up the HCD so that its read/write loop is driven by the controller was made based on past experience where similar (though older generation) Galil controllers would tend to crash (requiring a power cycle), if the loop were instead driven by state requests from the application at a high rate ($\gtrsim$50\,Hz). Though not analyzed here, the software could easily be modified to accommodate such a scenario for testing purposes. The \textbf{Controller I/O} actor could simply send state requests to the controller, rather than blocking as it awaits asynchronous updates. The result would be a very similar read/write loop sequence of messages (blue lines in Figs.~\ref{fig:actors} and \ref{fig:seq}).

\subsection{Results}

All of the tests in this section were based on 48-hr runs, with the system configuration summarized in Table~\ref{tab:system}. Note that the kernel had the real-time patch applied, though no effort was made to improve jitter [e.g., by assigning real-time priority to processes, core isolation, or the use of Non-Uniform Memory Access (NUMA) regions].

\begin{table}[ht]
\caption{Initial test system configuration.}
\label{tab:system}
\begin{center}
\begin{tabular}{|l|l|} %% this creates two columns
%% |l|l| to left justify each column entry
%% |c|c| to center each column entry
%% use of \rule[]{}{} below opens up each row
\hline
%\rule[-1ex]{0pt}{3.5ex}  Article title & 16 pt., bold, centered  \\
Component & Details \\
\hline
%\rule[-1ex]{0pt}{3.5ex}  Author names and affiliations & 12 pt., normal, centered   \\
Server & Dell PowerEdge 1950 \\
       & $4\times2.0$\,GHz Intel Xeon L5335 CPUs, 8\,GB RAM \\
OS     & CentOS 7.2, Kernel: 3.10.93-rt101\#1 SMP PREEMPT RT \\
Java   & Version 1.8.0\_66 \\
       & \texttt{UseParallelGC=true} \\
       & \texttt{InitialHeapSize=132120576} \\
       & \texttt{MaxHeapSize=2092957696} \\
CSW    & Prototype Version 0.4 \\
Akka   & Version 2.4.16 \\
Redis  & Version 3.2.6 \\
Galil Controller &  DMC 4080 Rev 1.1i, 9568 \\
                 &  TCP socket communications\\
Network Switch & SMCGS24C-SMART 1GbE \\
\hline
\end{tabular}
\end{center}
\end{table}

\subsubsection{Position accuracy}

\begin{figure}[ht]
\begin{center}
\includegraphics[width=0.8\linewidth]{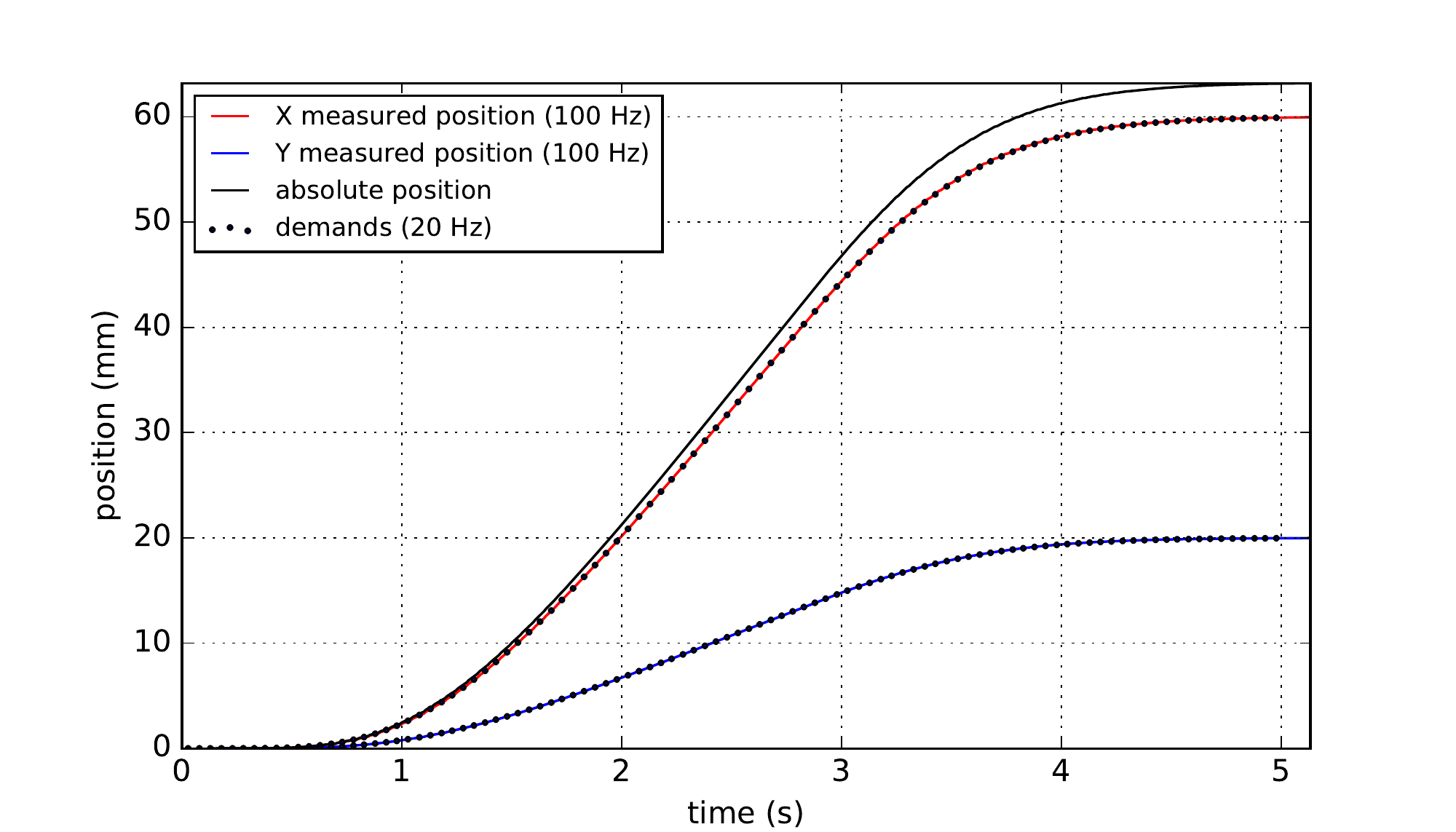}
\end{center}
\caption{\label{fig:hcd_demands}
Motion profile requested by the TCS at 20\,Hz (dots), and resulting simulated $x$- and $y$-positions measured by the motion controller at 100\,Hz (colored lines), as well as the absolute on-sky position (black line).}
\end{figure}

\begin{figure}[ht]
\begin{center}
\includegraphics[width=0.8\linewidth]{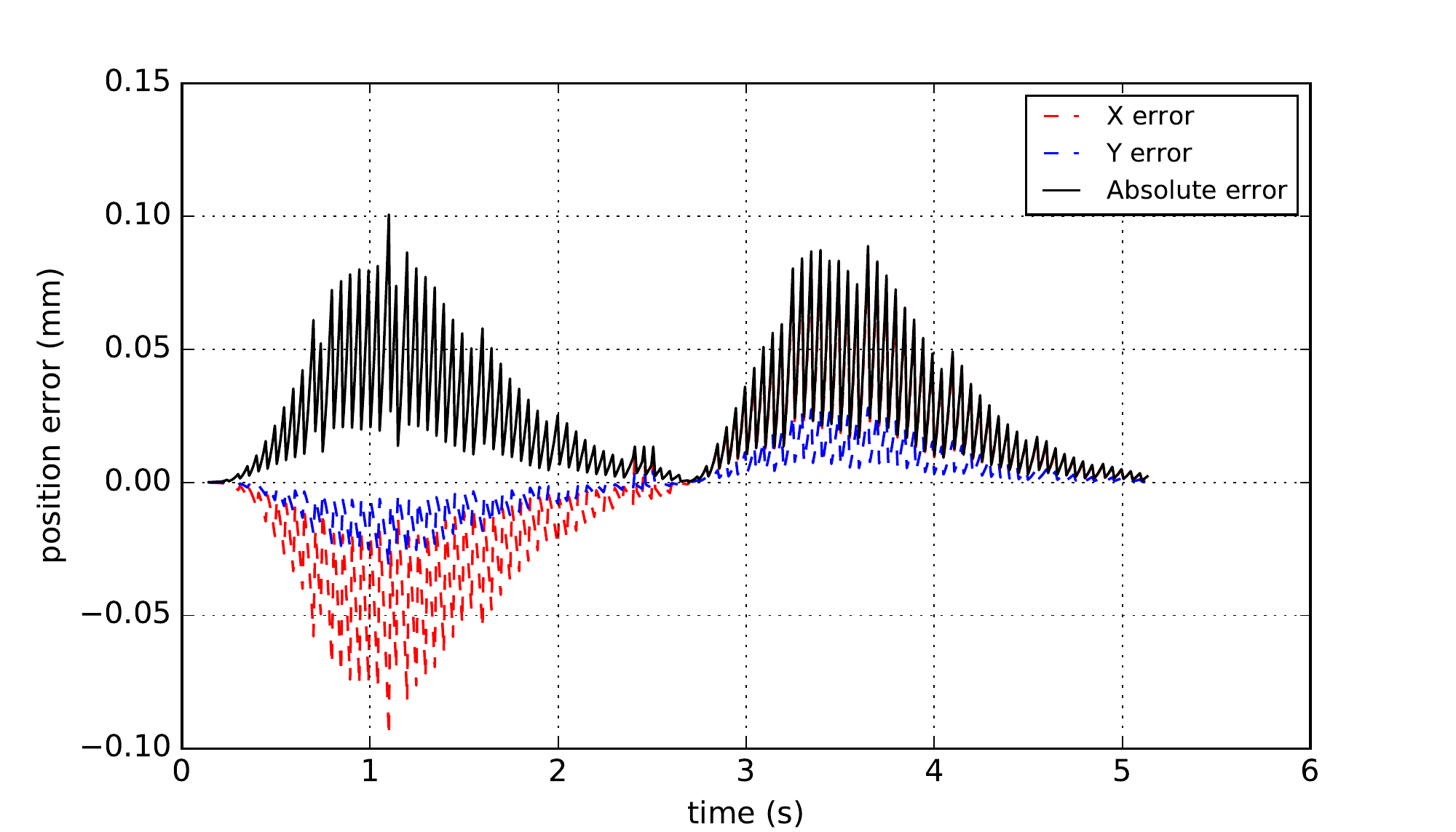}
\end{center}
\caption{\label{fig:hcd_demands_err}
Error between measured positions, and underlying ideal curve from which the TCS demands in Fig.~\ref{fig:hcd_demands} were sampled.}
\end{figure}

First, a single repetition of the ``S'' curve demands as published by the TCS (dots) are compared with the motor controller positions reported to the \textbf{Controller I/O} actor (solid lines) as part of the read/write loop in Fig.~\ref{fig:hcd_demands}. The error signal (evaluated at 100\,Hz as part of the read/write loop) is shown in Fig.~\ref{fig:hcd_demands_err}.

\begin{figure}[ht]
\begin{center}
\includegraphics[width=0.8\linewidth]{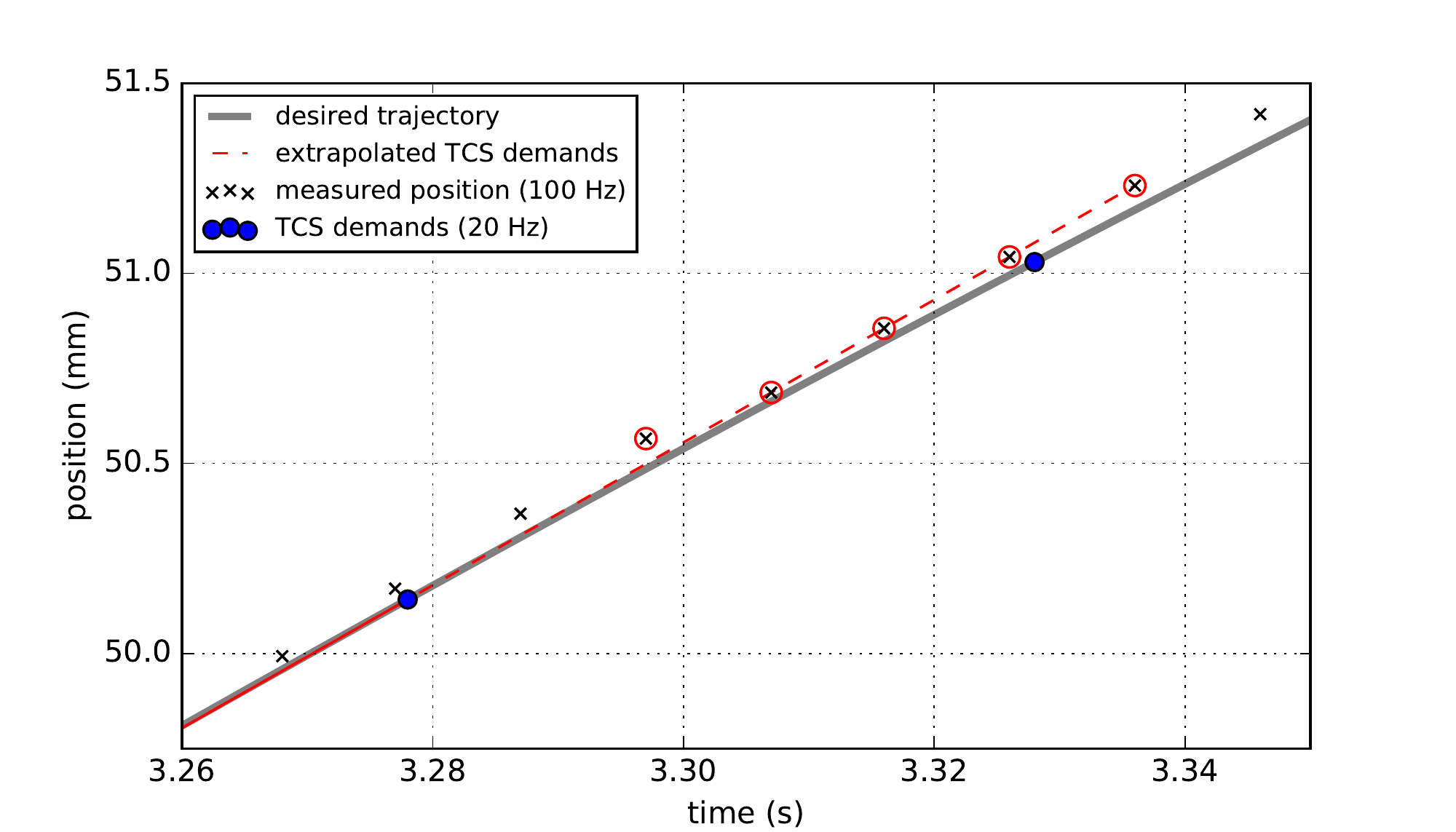}
\end{center}
\caption{\label{fig:hcd_demands_detail}
Detail showing demands, measurements, and extrapolation at the portion of the demand stream with greatest curvature. The red-circled `x' measurements result from controller demands derived from a linear extrapolation from the two TCS demands that arrived prior to 3.28\,s (solid red line shows where the fit occured, and the dashed red line the extrapolation).}
\end{figure}

The error is qualitatively similar to the ideal simulation of Fig.~\ref{fig:dither_extrap_err}, with the largest amplitude occurring during periods of maximum acceleration near 1.1\,s and 3.3\,s. The error amplitude is, however, somewhat smaller. Fig.~\ref{fig:hcd_demands_detail} shows a zoom-in of the latter portion of the curve, illustrating the linear extrapolation strategy.

This figure shows that the TCS demands are typically used by the HCD approximately two controller read/write cycles later, or a total lag of 20\,ms. This lag is substantially less than the 50\,ms assumed in Section~\ref{sec:tcs_demands}, explaining the improved error.

\subsubsection{Latency and jitter}

\begin{figure}[ht]
\begin{center}
\includegraphics[width=0.8\linewidth]{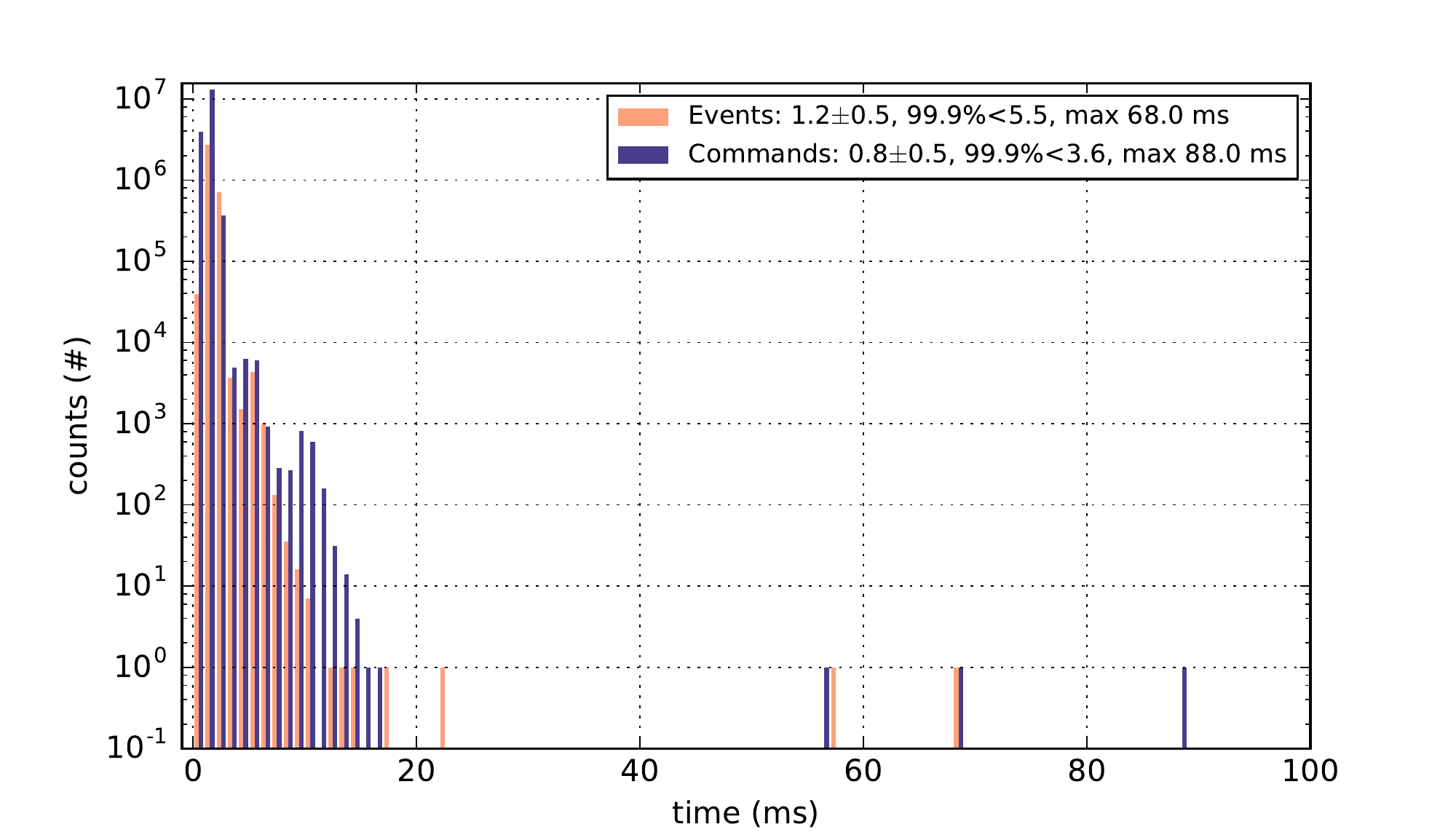}
\end{center}
\caption{\label{fig:events_commands}
Log histogram of CSW Event [interval (1)] and Command Service [interval (2)] message lags over 48\,Hr, with 1\,ms bins.}
\end{figure}

As the tests ran, timing statistics were collected through the use of \texttt{java.time.Instant.now().toEpochMilli()} calls to measure several key intervals, as indicated in Figs.~\ref{fig:actors} and \ref{fig:seq}. Interval (1) includes the transit time of events from the TCS to the POA assembly via the event service, and the conversion to a local Akka message by the \textbf{Event Subscriber} for the \textbf{Director}. Interval (2) measures the transit time of demands send by the POA Assembly to the Controller HCD via the Command Service (a thin wrapper over Akka messages sent remotely). Histograms of these timings are shown in Fig.~\ref{fig:events_commands}, with small horizontal offsets so that the bins do not overlap. The Event Service histogram contains 3456000 samples (20\,Hz over 48\,hr) and the Command Service (HCD) histogram 17280000 samples (100\,Hz over 48\,hr)

In both cases the mean latency is approximately 1\,ms (the resolution of the timing function used), with 99.9\% of the deliveries occuring in 5.5\,ms, and 3.6\,ms, for the Event and Command Services, respectively. Note that there are a handful of outliers ($>$20\,ms) during the 48\,hr experiment. If an event or demand is delayed by more than one full period ($>$50\,ms and $>$10\,ms, respectively), the relevant actor mailboxes will simply queue the messages. Then, when the actors become responsive again, they will rapidly process those queued messages until they are once again empty.

\begin{figure}[ht]
\begin{center}
\includegraphics[width=0.8\linewidth]{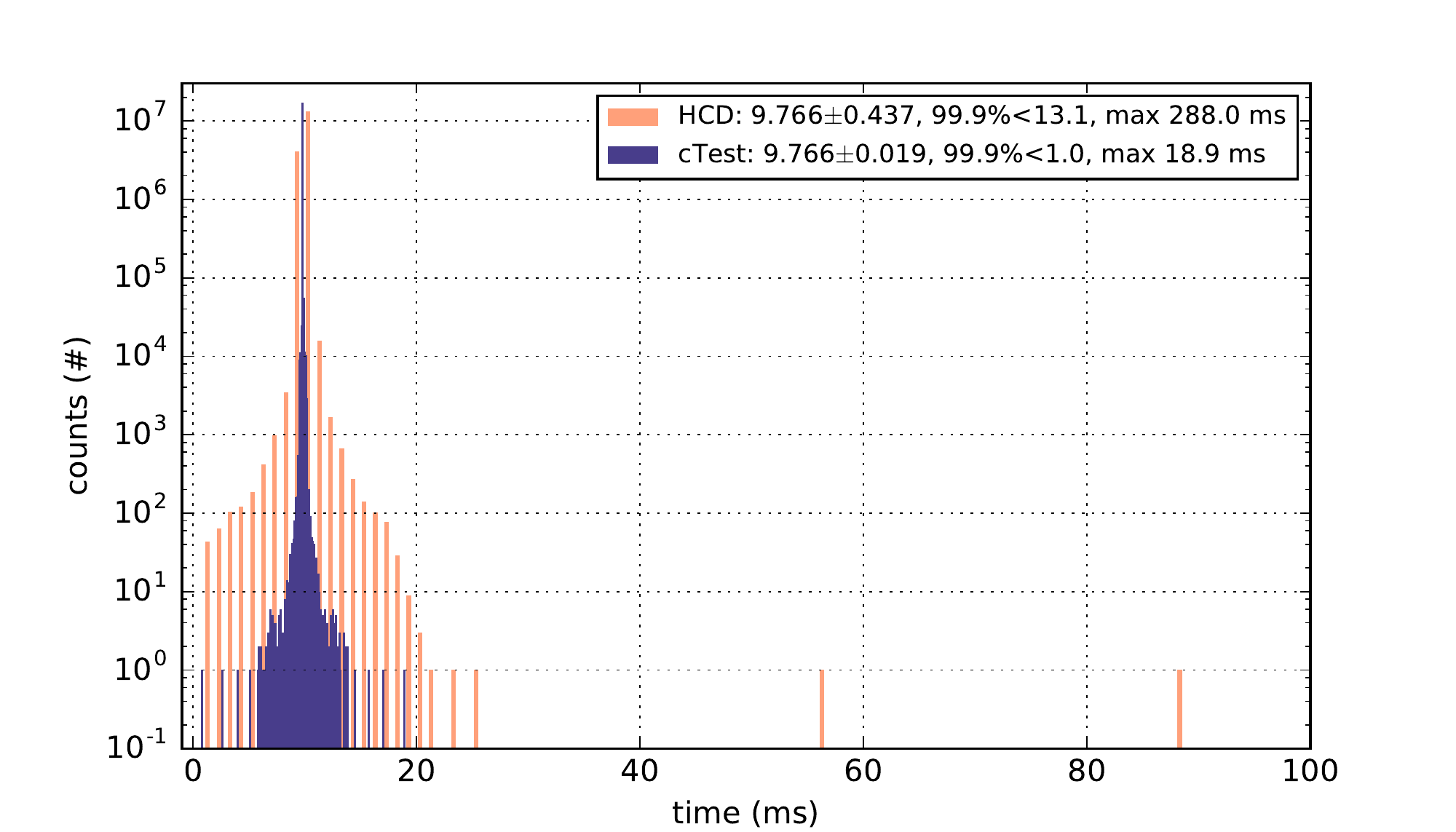}
\end{center}
\caption{\label{fig:io_period}
Log histogram of periods between state updates [interval (3)] received by HCD Controller I/O actor (1\,ms bins), as compared with a reference application written in C (0.1\,ms bins). While the $x$-axis is truncated to 100\,ms, there was a single measurement at 288\,ms.}
\end{figure}

Next, Fig.~\ref{fig:io_period} shows a histogram of the periods between state records reported by the motion controller (17280000 samples), as measured by the \textbf{Controller I/O} actor (orange). Note that this histogram is largely symmetric (unlike the previous histograms), since a long period will typically be followed by a comparably short period as it gets caught up. A reference application was written in C, cTest, that executes the same Galil commands (initiate asynchronous record pushes from the controller, and repeatedly calls \texttt{gRecord()} to retrieve them, followed immediately by new axis demand calls each time it returns), without any of the overheads of the JVM (blue). The two distributions have matching means, though the width produced by cTest is substantially smaller (standard deviation of 0.019\,ms vs. 0.437\,ms), and has none of the $>$20\,ms outliers that are evident for the Java application. The results of cTest demonstrate that the combined jitter inherent to the controller itself, and the OS, is no greater than $\sim$20\,ms.

If there is significant jitter in the JVM, taking longer than an expected I/O period, the Controller I/O actor will rapidly cycle through and report any queued records that were pushed by the controller during the pause.

Of course, some jitter in this control system is acceptable, since we showed that TCS demands can be used up to 50\,ms late; this seems like a good trade-off given the relative ease of development and portability of Java applications. The outliers we note in the Java case are also incredibly rare; in a 48\,hr period there were only 7 controller update loops that took longer than 20\,ms (out of 17280000). While one might be tempted to ignore them altogether, it is worth keeping such outliers in mind when considering the overall telescope system performance, and any potential hazards or loss of observing efficiency. For the example at hand, these outliers would likely only impact observing efficiency if they were to occur precisely during periods of large accelerations in a closed-loop AO dither (the WFSs would lose lock, and the system would need to re-acquire the sources, and possibly re-start a science observation). It is unlikely that any damage would occur, though there is a small possibility of a collision between the OIWFS Probes if significant jitter were to occur during a complex re-configuration\cite{chapin2016}.

On the other hand, the full suite of observatory software will have many more opportunities for lag and jitter to affect the overall system performance. It is therefore prudent to understand what causes the jitter, and to determine if there are any easy mitigations.

\subsubsection{Garbage collector tuning}

After some minimal investigation, it was found that the slow timing measurements from the previous section correlate well with the activities of the JVM garbage collector (GC). The basic operation of the GC is as follows: dynamic memory used throughout the life of an application comes from the ``heap'', whose size is defined on startup. As objects are created and destroyed, reference counting is used to determine whether memory is still in use. Periodically, the GC will check these counters, and returns unreferenced memory to the heap. Some GCs will also perform ``compacting'', which moves the memory still in use into contiguous blocks, leaving larger free blocks of memory for future allocation. Garbage collection temporarily halts code execution to freeze the state of the memory. To improve performance, GCs generally divide the heap into ``generations'', with young heap generations being checked more frequently, and eventually moving objects that have survived multiple young heap garbage collection events (or ``minor'' GC events) into older generations. The older generation of the heap is typically larger, and is checked less frequently (``major'' GC events).

\begin{figure}[ht]
\begin{center}
\includegraphics[width=0.8\linewidth]{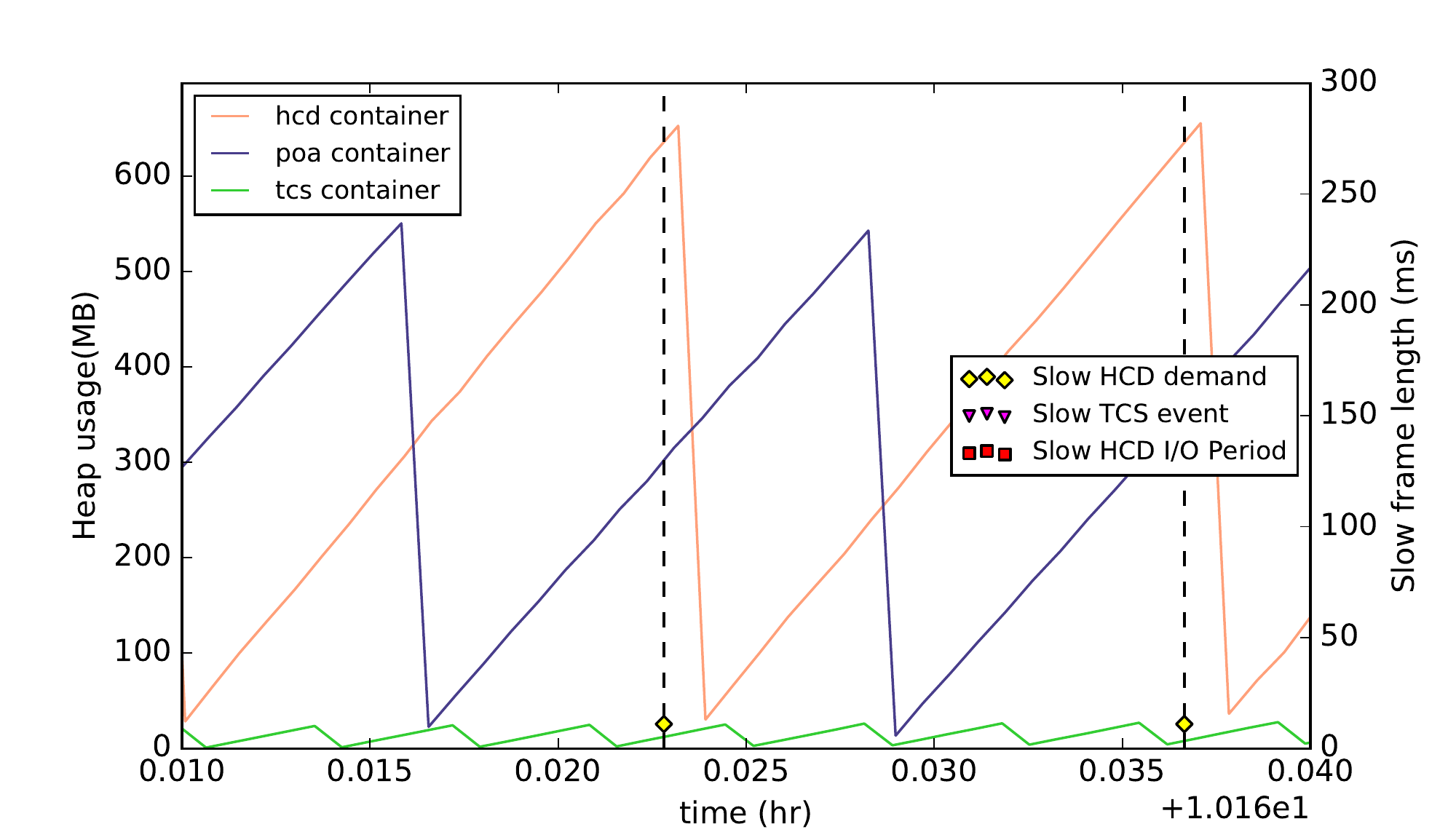}
\end{center}
\caption{\label{fig:young_heap_zoom}
Young heap memory usage detail. Abrupt drops are caused by minor GC events. Slow intervals have the following thresholds: 20\,ms for events, 10\,ms for demands, and 20\,ms for I/O periods.}
\end{figure}

\begin{figure}[ht]
\begin{center}
\includegraphics[width=0.8\linewidth]{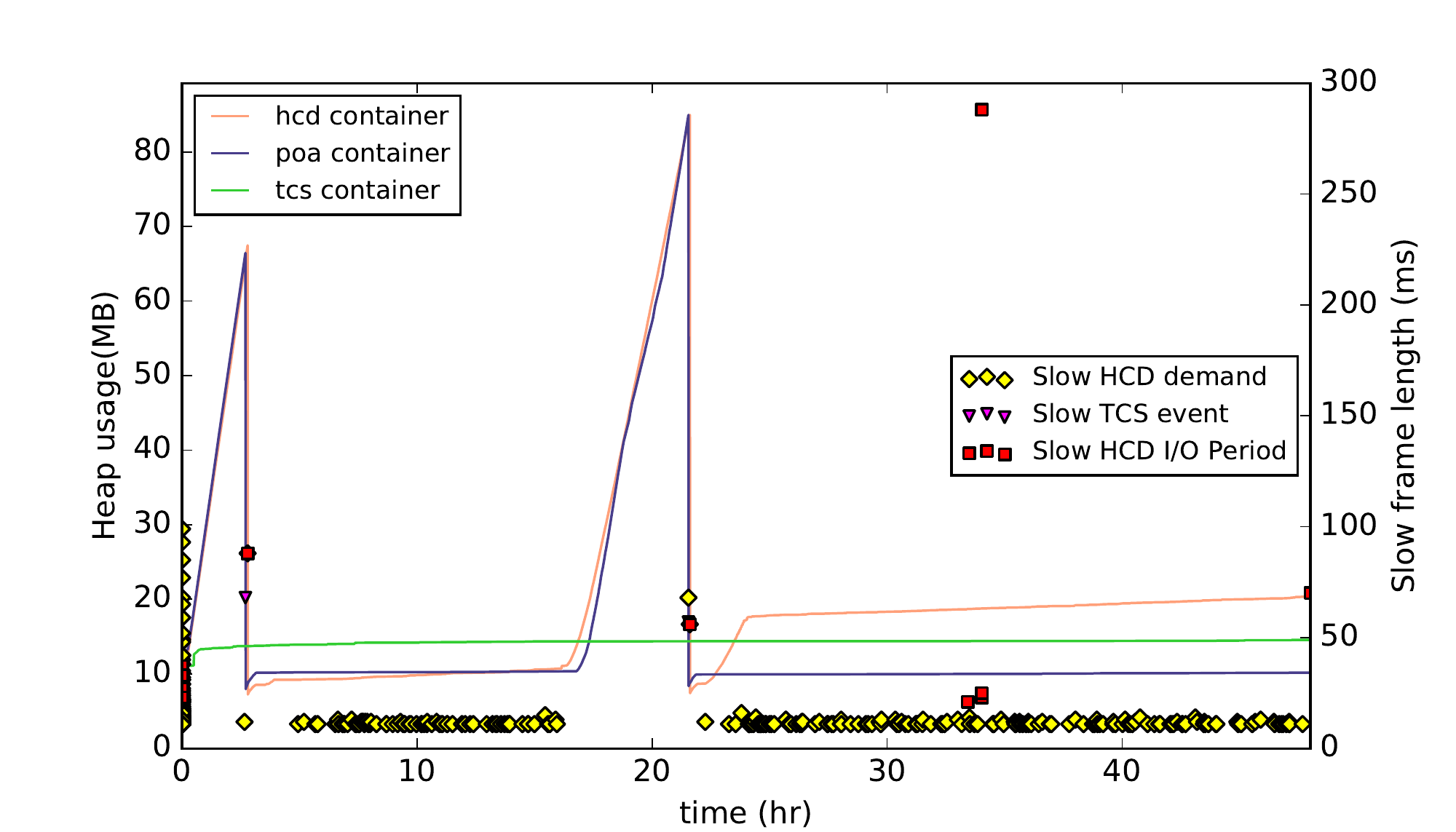}
\end{center}
\caption{\label{fig:old_heap}
Old heap memory usage (symbols have same meanings as Fig.~\ref{fig:young_heap_zoom}).}
\end{figure}

While running the three test applications, the states of their GCs were probed at a frequency of 1\,Hz using \texttt{jstat -gc}. The applications were also modified so that they would log unexpectedly slow intervals: (1) events that took longer than 20\,ms to reach the Assembly \textbf{Director} from the TCS; (2) commands (demands) that took longer than 10\,ms to reach the HCD \textbf{Director} from the Assembly \textbf{Director}; and (3) \textbf{Controller I/O} read/write loop periods that took longer than 20\,ms (i.e., more than 10\,ms late, as each period is normally 10\,ms). Fig.~\ref{fig:young_heap_zoom} shows a portion of the young generation heap usage over 0.03\,hr (108\,s), illustrating the growth in heap usage as objects are created, followed by abrupt drops when minor GC events occur. It also shows two examples where slow HCD demands correspond to minor GC events in the HCD container JVM. In general, it is found that the tail of slow demands apparent in Fig.~\ref{fig:events_commands} at 10--20\,ms correspond to minor GC events. While the threshold for slow events and I/O periods was set to a longer 20\,ms period in this experiment (and therefore do not appear as often in the heap usage plots), brief tests with shorter logging thresholds confirm that the equivalent tails in the events and HCD period histograms (Figs.~\ref{fig:events_commands} and \ref{fig:io_period}) can also generally be explained by the minor GC events.

Next, the old heap is shown for the entire 48-hr run in Fig.~\ref{fig:old_heap}. In addition to the tail of slow HCD demands at 10--20\,ms, a handful of much slower HCD demands and I/O periods are apparent (up to nearly 300\,ms), typically correlated with the major GC events, as well as a cluster of slow demands/events/periods at the start of the run. The initial cluster is likely due to some initial automated tuning of the JVM Just-In-Time (JIT) optimizations and/or GC, and is ignored as part of a ``warmup interval'' when creating the histograms in Figs.~\ref{fig:events_commands} and \ref{fig:io_period}. It is unclear why the old generation heap usage grows slowly for a long period after a major GC event (followed by a faster rise), nor is it understood what caused the particularly slow I/O periods at $\sim$34\,hr, as they do not appear to correlate with the GC.

%\item Takashi wonders about two big peaks in Fig.~\ref{fig:old_heap}. \textbf{ the old generation was consumed for some reason. I also don't know why, but I suspect there was a problem with the connection with the motion controller and your benchmark program got confused. This is because while the heap usage is increasing, there was no "Slow HCD demand" recorded, which is recorded from time to time in other period.}

%\begin{figure}[ht]
%\begin{center}
%\includegraphics[width=\linewidth]{hwMultiContainerjvm_heap_young20.pdf}
%\end{center}
%\caption{\label{fig:young_heap}
%Correlation between outlier events (symbols) and state of the JVM young memory heap. Garbage collection events occur at the sudden drops.}
%\end{figure}

We then performed a simple experiment to see if basic GC tuning could improve the outlier performance. Noting that the Parallel Collector used in our initial tests (Table~\ref{tab:system}) is optimized for throughput over jitter, we tried the newer Garbage First Garbage Collector (G1GC), which theoretically suffers less jitter (and which has since become the default in Java 9). While there are many options with which to experiment\footnote{http://www.oracle.com/technetwork/articles/java/g1gc-1984535.html}, we opted only to set the following: \texttt{-XX:+UseG1GC -J-XX:MaxGCPauseMillis=10}. This second parameter is the target value for the maximum time allowed by the GC for garbage collection; we set it to 10\,ms (to match the shorter controller I/O loop period), down from the default of 200\,ms.

\begin{figure}[ht]
\begin{center}
\includegraphics[width=0.8\linewidth]{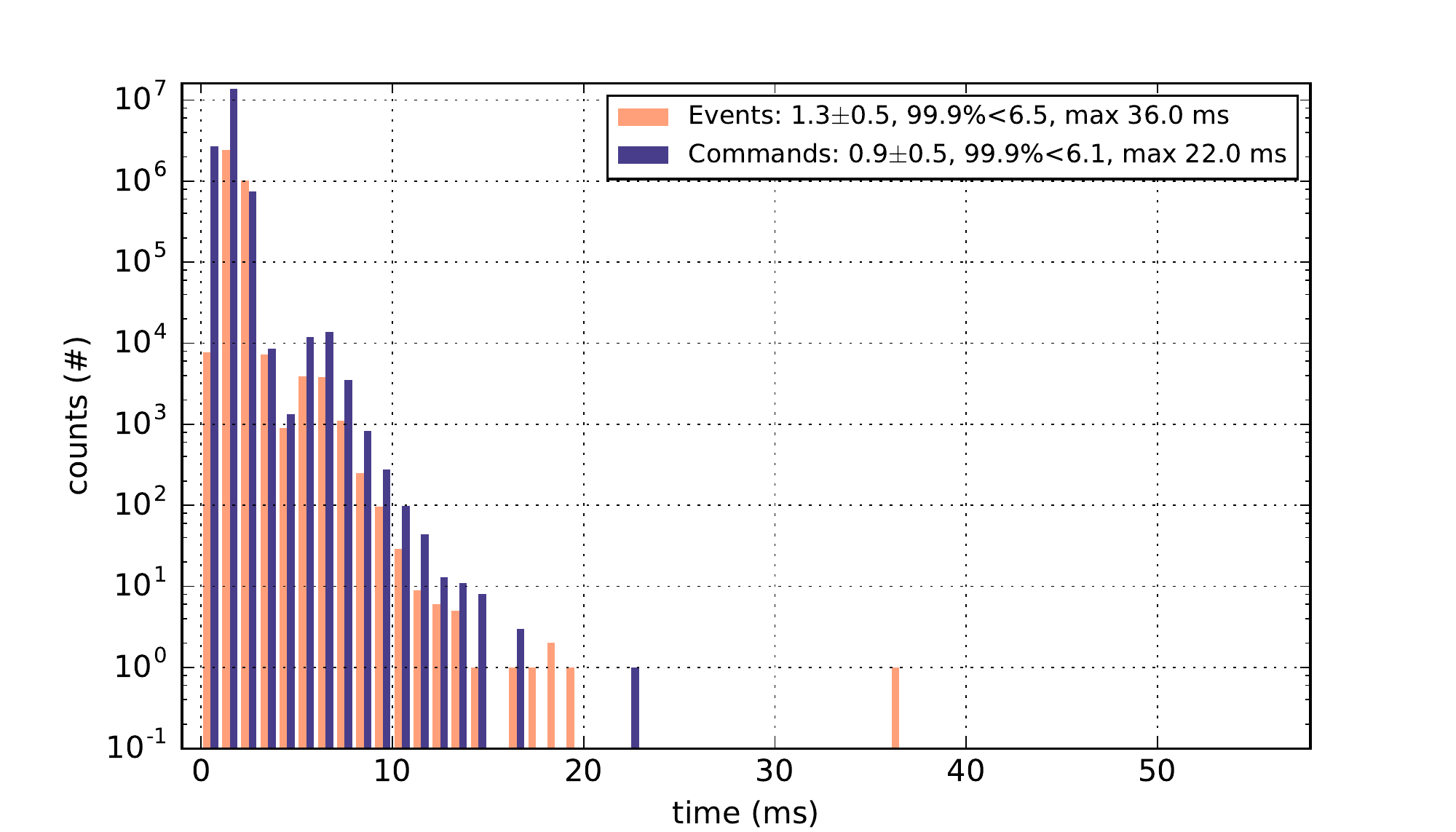}
\end{center}
\caption{\label{fig:events_commands_g1gc}
Log histogram of CSW Event and Command Service message lags over 48\,hr, now using the G1GC garbage collector, with a target maximum GC time of 10\,ms (compare with Fig.~\ref{fig:events_commands}).
}
\end{figure}

\begin{figure}[ht]
\begin{center}
\includegraphics[width=0.8\linewidth]{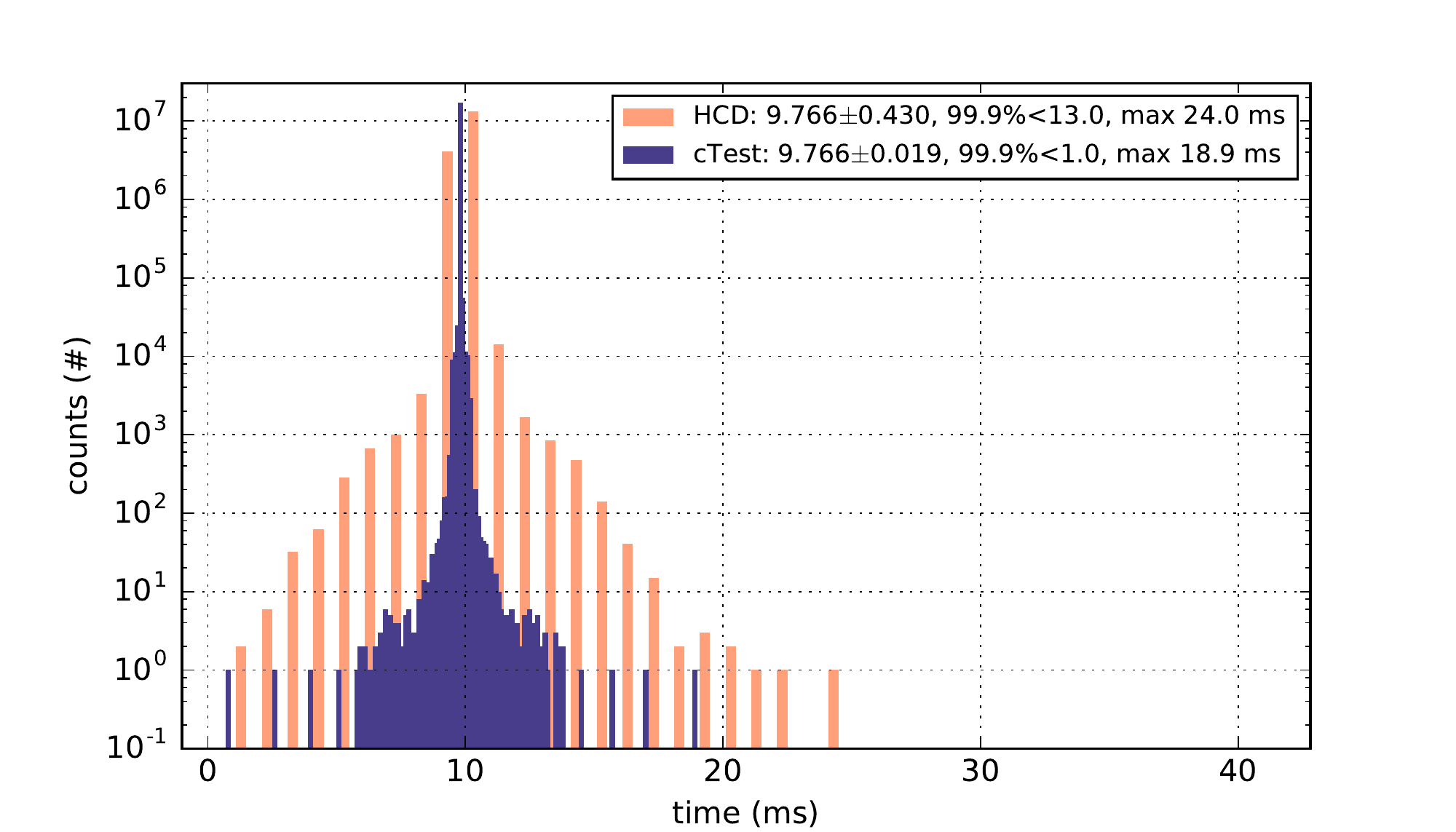}
\end{center}
\caption{\label{fig:io_period_g1gc}
Log histogram of periods between state updates received by HCD Controller I/O actor (1\,ms bins), as compared with a reference application written in C (0.1\,ms bins), now using the G1GC garbage collector, with a target maximum GC time of 10\,ms (compare with Fig.~\ref{fig:io_period}).
}
\end{figure}

Figs.~\ref{fig:events_commands_g1gc} and \ref{fig:io_period_g1gc} show timing histograms using data collected over a 48\,hr run, and are directly comparable to Figs.~\ref{fig:events_commands} and \ref{fig:io_period} for the Parallel GC. Note that the large (though infrequent) outliers have been completely removed; other than a single TCS event that arrived after 36\,ms, all other measured intervals were within 25\,ms.

%\begin{figure}[ht]
%\begin{center}
%\includegraphics[width=\linewidth]{hwMultiContainerG1GCjvm_heap_young20.pdf}
%\end{center}
%\caption{\label{fig:old_heap_g1gc}
%Correlation between outlier events (symbols) and state of the JVM young memory heap using G1GC and 10\,ms max GC time. Major garbage collection events occur at the sudden drops.}
%\end{figure}

\begin{figure}[ht]
\begin{center}
\includegraphics[width=0.8\linewidth]{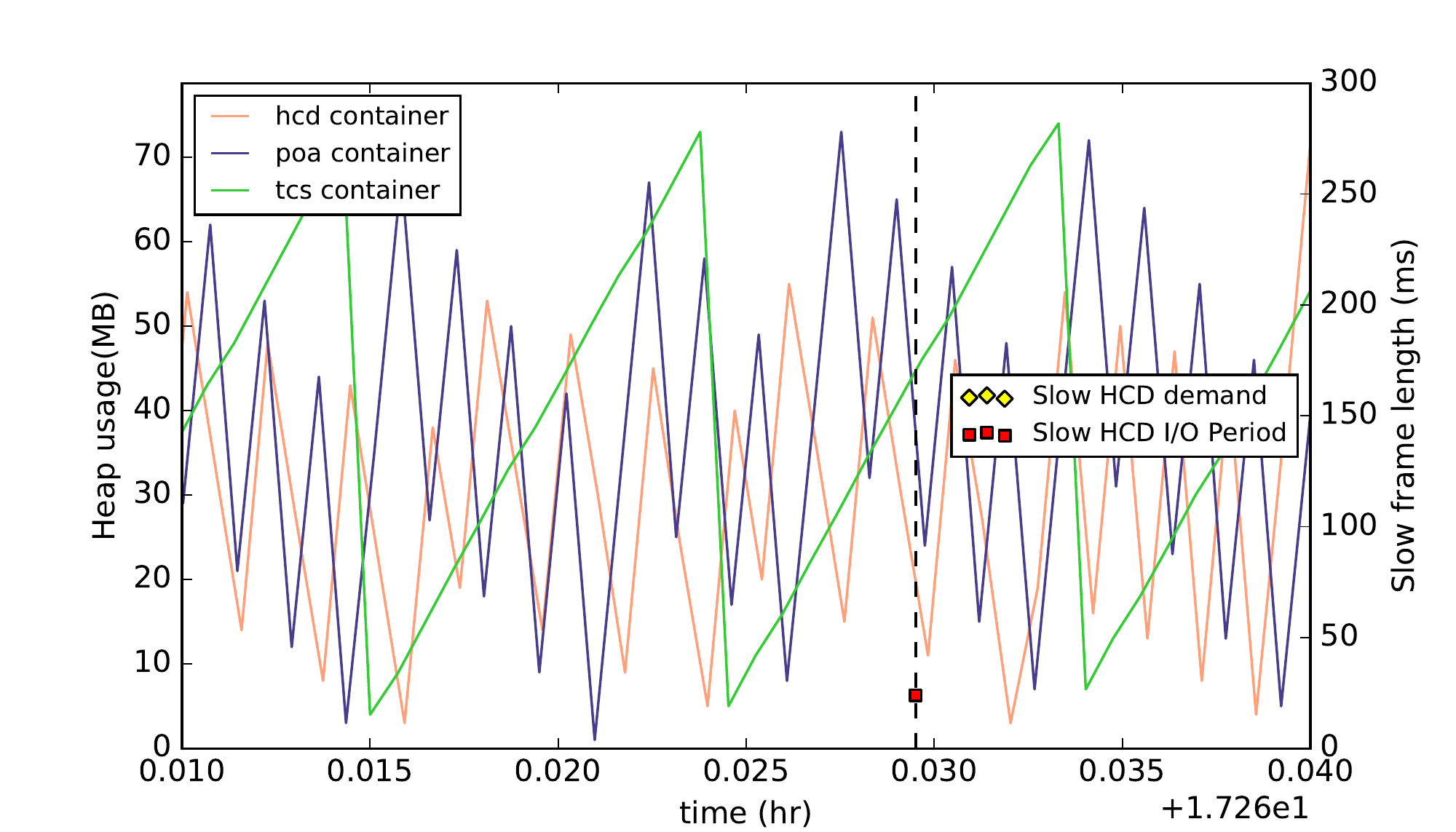}
\end{center}
\caption{\label{fig:young_heap_g1gc_zoom}
Young heap memory usage detail when using the G1GC garbage collector (compare with Fig.~\ref{fig:young_heap_zoom}).}
\end{figure}

\begin{figure}[ht]
\begin{center}
\includegraphics[width=0.8\linewidth]{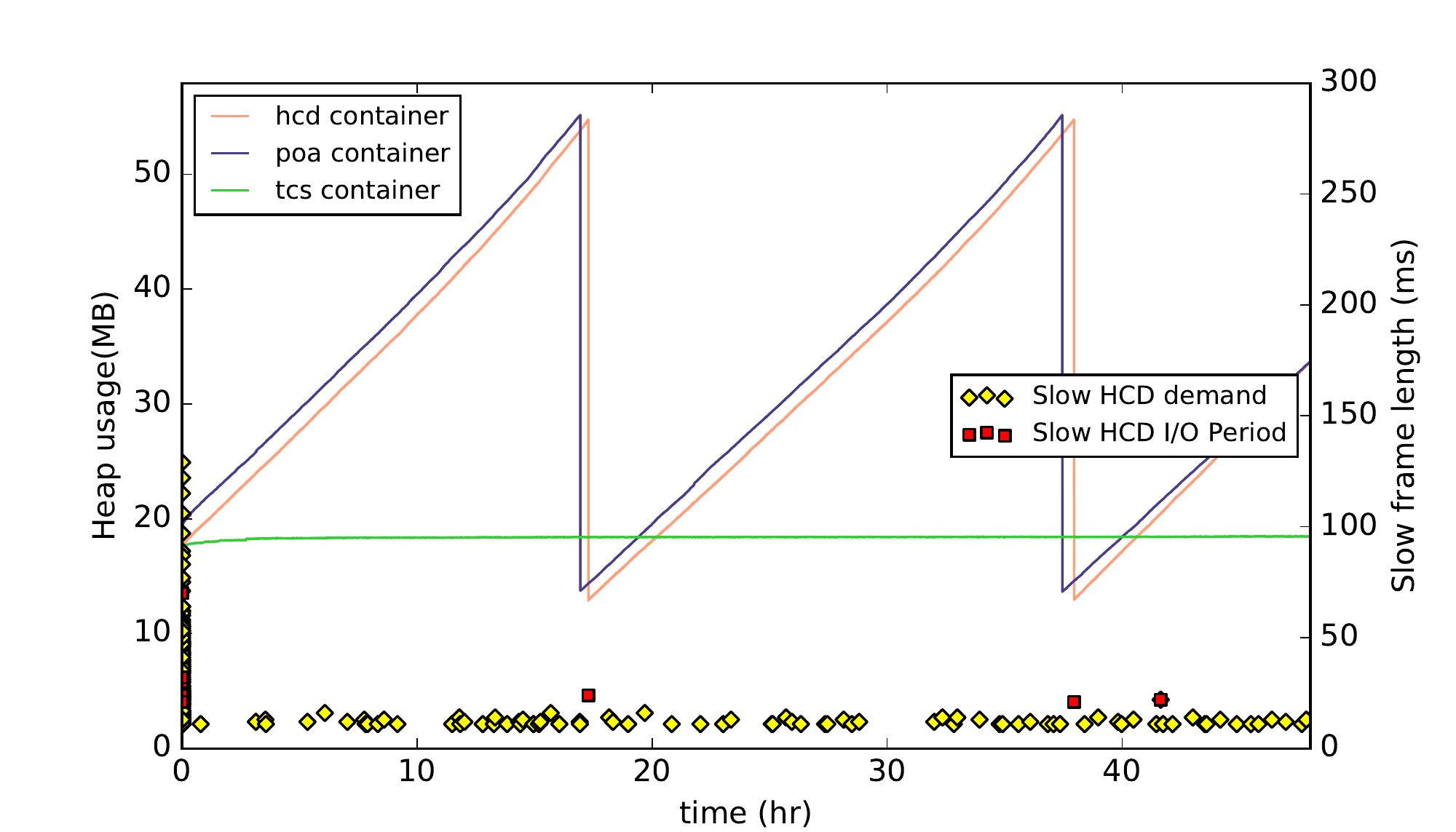}
\end{center}
\caption{\label{fig:old_heap_g1gc}
Old heap memory usage when using the G1GC garbage collector (compare with Fig.~\ref{fig:old_heap}).
}
\end{figure}

Next, Figs.~\ref{fig:young_heap_g1gc_zoom} and \ref{fig:old_heap_g1gc} show the young and old heap usage plots, respectively, for comparison with Figs.~\ref{fig:young_heap_zoom} and \ref{fig:old_heap} for the Parallel GC. Note, in particular, that minor GC events are now much more frequent, which is consistent with the expectation that it is spreading the work more evenly over time, and taking fewer large pauses. Like the earlier test with the Parallel GC, two major GC events occur over the 48\,hr test period, though the pauses only appear to correlate with ``Slow HCD I/O'' periods of up to $\sim$25\,ms (vs. $\sim$100\,ms for the earlier tests, excluding the unexplained pause of $\sim$300\,ms in Fig.~\ref{fig:io_period}). In general, the behaviour of both the young and old generations of the heap appear to be more predictable. Finally, there is again a cluster of slow measurements on startup that have been ignored in the histograms as part of the warmup period.

These results show that rare events of significant jitter in the JVM ($\gtrsim$20\,ms) can be effectively mitigated with minimal tuning of the GC. Furthermore, this improvement does not appear to come at the expense of increased jitter on shorter timescales; the standard deviations of all measured intervals are essentially identical in the two experiments that were performed. Further GC experimentation may yield even better performance, although it is clear that we are already approaching the expected limit for outlier performance, as the slowest measured I/O loop periods in our Java test application ($\sim$25\,ms) only slightly exceed the longest periods measured in our reference C application ($\sim$18\,ms), as shown in the yellow and blue histograms of Fig.~\ref{fig:io_period_g1gc}, respectively.

\section{Summary and future work}

In this paper we have demonstrated the feasibility of closed-loop AO dithering with IRIS and NFIRAOS from the perspective of their slow opto-mechanical mechanisms. Taking an approach in which the telescope mount and wavefront sensor positioners within IRIS and NFIRAOS are driven with a smooth ``S'' profile produced by the TCS at 20\,Hz, the expected position uncertainties are within 0.2\,mm in the focal plane. This limit was established to avoid saturating the NFIRAOS tip/tilt stage (which compensates for pointing errors in real-time). Key elements of the control system for one of the positioners were prototyped in Java, including the propagation of demands using Common Software Services supplied by TMT, and interaction with a real Galil 8-axis Motion Controller. An essential component of our control strategy is linear extrapolation of past TCS demands by the low-level Hardware Control Daemon (HCD) evaluated at the precise instants they are required. Finally, we study jitter in the system, finding that significant ($\gtrsim$20\,ms) pauses can generally be associated with activities of the Java garbage collector (GC). Switching from the default settings used in Java 8 (Parallel garbage collector), to the newer Garbage First Garbage Collector (G1GC), with the inclusion of the \texttt{MaxGCPauseMillis} parameter set to 10\,ms, removes essentially all of the significant pauses measured over 48\,hr. While further tuning may continue to improve the jitter performance, we note that the greatest outliers in our Java test applications only slightly exceed those of a reference application written in C that does not use a GC.

A number of future tests are planned. First, the HCD may be updated to either: (a) drive a real motor; or (b) include an improved simulation, with a real PID loop, and at least trapezoidal velocity profiles (as the current simulation allows instantaneous velocity changes). The software components and CSW services should be run on separate physical servers to include the effects of the network. Real-time performance can be improved by setting real-time priorities, core isolation, and the use of NUMA regions; in particular, it would be interesting to see if the width of the reference C application control loop period histogram can be reduced. If it is possible to significantly improve the C application performance, it may then be worth further experimentation with the many GC tuning parameters. Finally, once the production TMT CSW services are released, these test applications should be re-written to make use of them, and re-benchmarked.

\acknowledgments

The TMT Project gratefully acknowledges the support of the TMT collaborating institutions.  They are the California Institute of Technology, the University of California, the National Astronomical Observatory of Japan, the National Astronomical Observatories of China and their consortium partners, the Department of Science and Technology of India and their supported institutes, and the National Research Council of Canada.  This work was supported as well by the Gordon and Betty Moore Foundation, the Canada Foundation for Innovation, the Ontario Ministry of Research and Innovation, the Natural Sciences and Engineering Research Council of Canada, the British Columbia Knowledge Development Fund, the Association of Canadian Universities for Research in Astronomy (ACURA) , the Association of Universities for Research in Astronomy (AURA), the U.S. National Science Foundation, the National Institutes of Natural Sciences of Japan, and the Department of Atomic Energy of India.

% References
\bibliography{refs} % bibliography data in refs.bib
\bibliographystyle{spiebib} % makes bibtex use spiebib.bst

\end{document}